\documentclass[onecolumn,showpacs,preprintnumbers,elsart]{revtex4}
\usepackage{amsmath}
\usepackage{mathrsfs}
\usepackage{graphicx}
\usepackage{dcolumn}
\usepackage{bm}
\usepackage[center]{subfigure}
\usepackage{color}
\begin{document}

\title{Quasi-compactons in inverted nonlinear photonic crystals}
\author{Yongyao Li$^{1,2}$}
\email{yongyaoli@gmail.com}
\author{Boris A. Malomed$^{3,4}$}
\author{Jianxiong Wu$^{1}$}
\author{Wei Pang$^{5}$}
\author{Sicong Wang$^{2}$}
\author{Jianying Zhou$^{2}$}
\email{stszjy@mail.sysu.edu.cn}
\affiliation{$^{1}$Department of Applied Physics, South China Agricultural University,
Guangzhou 510642, China \\
$^{2}$State Key Laboratory of Optoelectronic Materials and Technologies,\\
Sun Yat-sen University, Guangzhou 510275, China\\
$^{3}$ Department of Physical Electronics, School of Electrical Engineering,
Faculty of Engineering, Tel Aviv University, Tel Aviv 69978, Israel\\
$^{4}$ ICFO-Institut de Ciencies Fotoniques, Mediterranean Technology
Park,08860 Castelldefels (Barcelona), Spain\\
$^{5}$ Department of Experiment Teaching, Guangdong University of
Technology, Guangzhou 510006, China.}

\begin{abstract}
We study large-amplitude one-dimensional solitary waves in photonic crystals
featuring competition between linear and nonlinear lattices, with minima of
the linear potential coinciding with maxima of the nonlinear \textit{%
pseudopotential}, and vice versa (\textit{inverted nonlinear photonic crystal%
}s, INPhCs), in the case of the saturable self-focusing
nonlinearity. Such crystals were recently fabricated using a mixture
of SU-8 and Rhodamine-B optical materials. By means of numerical
methods and analytical approximations, we find that large-amplitude
solitons are broad sharply localized stable pulses
(\textit{quasi-compactons}, QCs). With the increase of the total
power, $P$, the QC's centroid performs multiple switchings between
minima and maxima of the linear potential. Unlike cubic INPhCs, the
large-amplitude solitons are mobile in the medium with the saturable
nonlinearity. The threshold value of the kick necessary to set the
soliton in motion is found as a function of $P$. Collisions between
moving QCs are considered too
\end{abstract}

\maketitle






\section{Introduction and the model}

Photonic and matter waves propagating under the combined action of
linear and nonlinear lattice (LL and NL) potentials exhibit a plenty
of dynamics \cite{non-soliton}. In particular, solitons in such
media is a topic of considerable current interest, see recent review
\cite{RMP_Kar} and original works \cite{1D,Thaw} devoted to the
studies of one-dimensional (1D) solitons. Other works were dealing
with solitons \cite{2D,Abd} and localized vortices \cite{2Dvort} in
two-dimensional (2D) versions of such systems, which represent, in
particular, photonic crystal fibers \cite{PCF} and 2D photonic
crystals \cite{PC} made of nonlinear materials. The evolution of
solitons in these systems obeys the nonlinear Schr\"{o}dinger
equation (NLSE), in which the LL and NL are represented,
respectively, by a usual
periodic potential and by a periodic \emph{pseudopotential} \cite%
{LCQian,RMP_Kar}, which is induced by a spatially periodic
modulation of the local nonlinearity coefficient. In optics, the NL
represents the mismatch between the nonlinearity of the host
material and the stuff filling voids of the photonic-crystal-fibers
structure, which may be air, another solid material
\cite{all-solid}, or a liquid crystal \cite{LC}. The same model, in
the form of the Gross-Pitaevskii equation, applies to matter waves
in a Bose-Einstein condensate (BEC) which is trapped in a
combination of a linear periodic potential, created by an optical
\cite{OL} or magnetic \cite{magnetic} lattice, and a pseudopotential
lattice, that may be induced by a periodic
modulation of the local nonlinearity provided by a properly patterned \cite%
{RMP_Kar} external magnetic \cite{FR-magn} or optical fields \cite{FR-magn}.

Recently, a combination of \emph{competing} $\pi
$-out-of-phase-juxtaposed LL and NL, with maxima of the refractive
index coinciding with minima of the local strength of the
self-focusing nonlinearity and vice versa, was considered in several
works \cite{Kar12,LYY1,LYY2}. This medium may be naturally called an
\emph{inverted nonlinear photonic crystal} (INPhC). It was reported
that INPhCs could be fabricated by means of the technique based on
direct laser writing in silica \cite{Blomer}. It was also predicted
that a similar setting can be created in a virtual form, using the
electromagnetically induced transparency acting on dopant atoms
periodically distributed in a passive matrix \cite{LYY2}. Due to the
competition between the LL and NL, solitons in INPhCs may feature
specific power-dependent properties, such as double symmetry
breaking \cite{LYY2}. While previous studies of INPhCs were dealing
with the\emph{\ }Kerr (cubic) nonlinearity, in this work we consider
solitary waves in the system with saturable nonlinearity, which
occurs in various optical media. The extension of the analysis for
this nonlinearity is natural, whilst solitons in INPhCs feature the
strong sensitivity to the power.

\begin{figure}[tbp]
\centering
\subfigure[]{ \label{fig_1 a}
\includegraphics[scale=0.18]{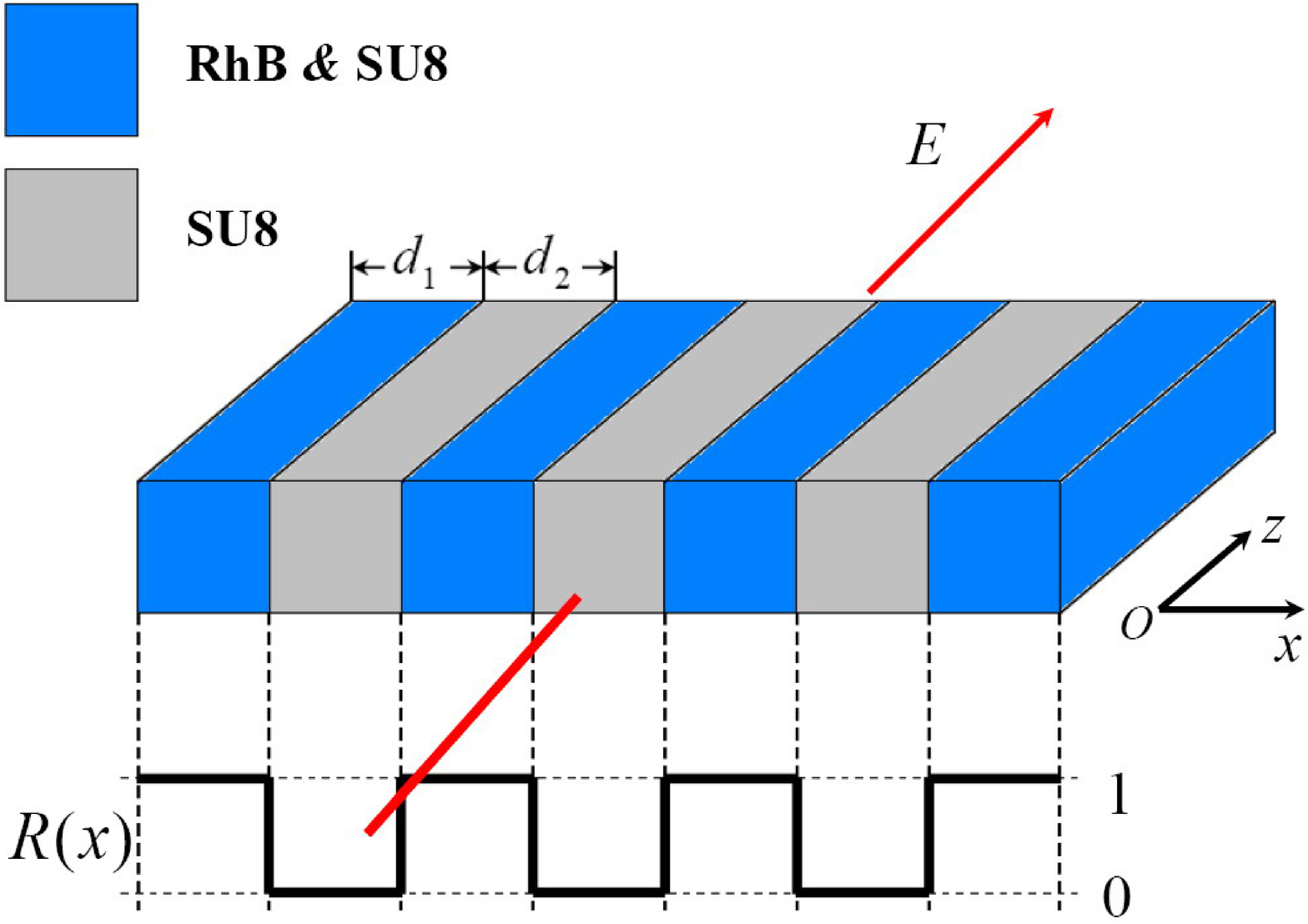}} \hspace{0.02in}
\subfigure[]{ \label{fig_1_b} \includegraphics[scale=0.16]{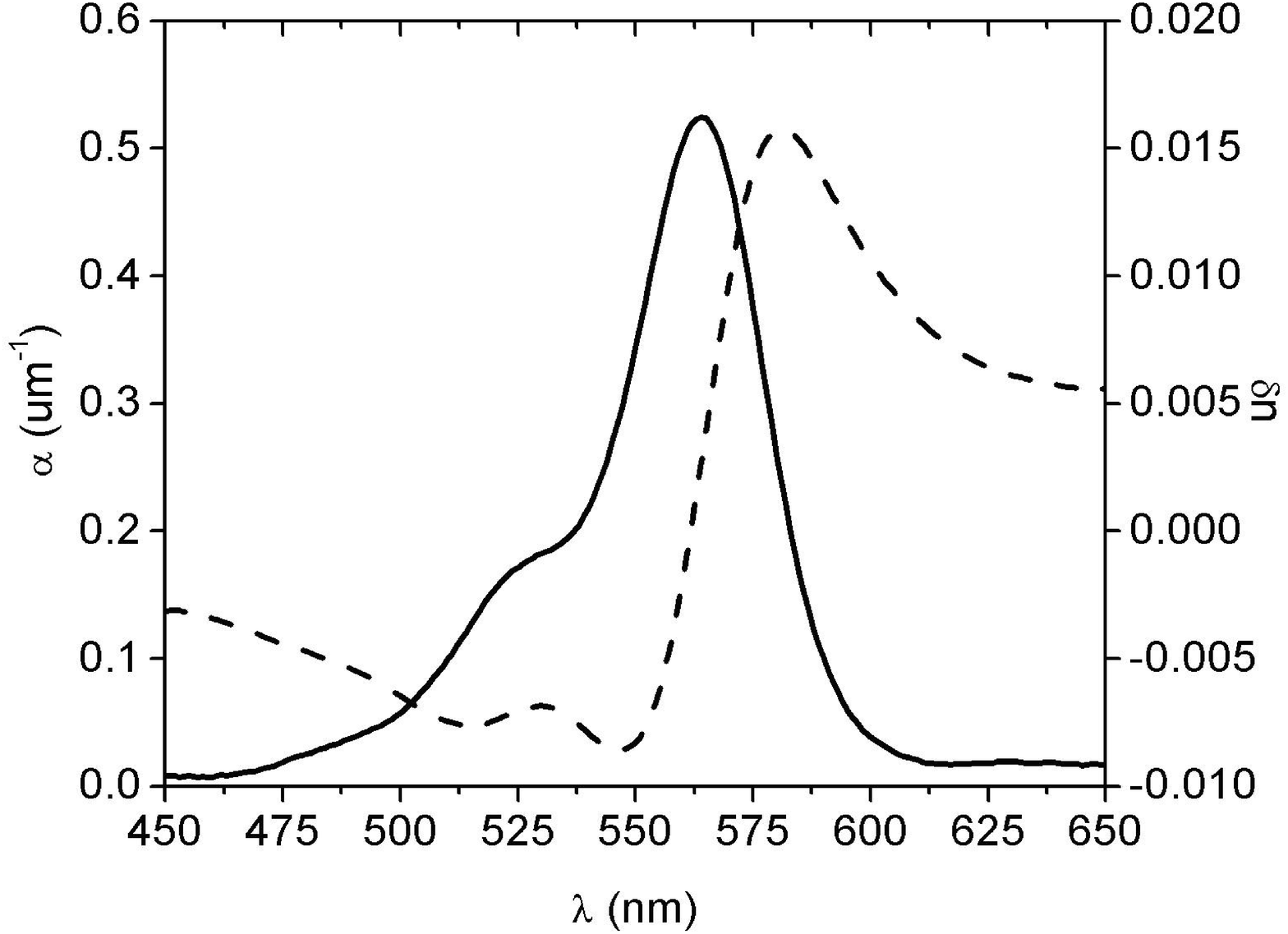}}
\caption{(Color online) (a)The one-dimensional nonlinear photonic
crystal, with the blue and gray areas depicting the nonlinear and
linear stripes, respectively. (b) The wavelength dependence of the
absorption coefficient and refractive-index variation,
$\protect\alpha $ and $\Delta n$ (solid and dashed lines,
respectively) in the SU-8 - RhB mixture.} \label{fig_1}
\end{figure}

First, we propose an experiment setup to realize an INPhC with a
saturable nonlinearity. Recent works reported the creation of
resonantly absorbing waveguide arrays and {\it imaginary-part
photonic crystals}, that feature a spatially periodic modulation of
the absorption coefficient, built on the basis of the SU-8 polymer
(a commonly used transparent negative photoresist) doped with
Rhodamine B (RhB, a dye featuring saturable absorption)
\cite{MNF,LJT}, see Fig. \ref{fig_1} for a schematic setup.

A similar pattern can be used for our purposes. In the paraxial
approximation, the light propagation in the array obeys the
spatial-domain NLSE for the local amplitude of the electromagnetic field, $%
u(z,x)$, where $z$ and $x$ are the propagation distance and transverse
coordinate:%
\begin{equation}
iu_{z}=-{\frac{1}{2k}}u_{xx}-{\frac{(k_{0}\delta n+i{\alpha /2})R(x)}{%
1+|u|^{2}}}u.  \label{RhBeq}
\end{equation}%
Here, $R(x)$ is the array's structure function which is to be taken
as per Fig. \ref{fig_1 a}, $k=k_{0}n$, where $n=1.62$ is the
refractive index of pure SU-8, $k_{0}=2\pi /\lambda $, and $\lambda
$ is the wavelength of light in vacuum. Further, $\alpha $ and
$\delta n$ are the absorption coefficient and \emph{refractive index
difference} between the SU-8 - RhB mixture and pure SU-8,
respectively, which are coupled by the Kramers-Kronig relation, and
are plotted versus $\lambda $ in Fig. \ref{fig_1_b} \cite{LJT}.
According to experimental data \cite{LJT}, in the relevant range of
far blue shift from the absorption peak ($\lambda \approx 460$ nm),
$\alpha /2\simeq (1/20)k_{0}\left\vert \delta n\right\vert $, which
allows one to neglect the dissipative term in Eq. (\ref{RhBeq}),
simplifying it to

\begin{equation}
iu_{z}=-{\frac{1}{2}}u_{xx}+{\frac{V(x)}{1+|u|^{2}}}u,  \label{NLSP}
\end{equation}%
where $k$ was removed by rescaling of $x$, and $V(x)\equiv -\left(
k_{0}\delta n\right) R(x).$

In the system under consideration, light experiences strong
saturable self-focusing induced by RhB in the mixed material, while
the nonlinearity of pure SU-8 is negligible. On the other hand, the
refractive index of the mixture is smaller than in pure SU-8, i.e.,
$\delta n<0$ at $\lambda =460$ nm, hence function $V(x)$ in Eq.
(\ref{NLSP}) takes positive values, and the waveguide array meets
the condition of the $\pi $ phase shift between the spatial
modulations on the linear and nonlinear local characteristics of the
medium, thus realizing the INPhC with the saturable nonlinearity.

The objective of this work is to study the existence, stability,
mobility, and interactions of solitons in the INPhC model based on
Eq. (\ref{NLSP}). In Sec. II, we study properties of the solitons by
means of numerical simulations and analytical approximations. It
will be demonstrated that they feature sharp localization, i.e., a
quasi-compacton (QC) shape. Following the increase of the total
power, $P$, the QC switches its position between maxima and minima
of the linear refractive index. In Sec. III, we study mobility of
the QCs, imposing the phase tilt onto them, i.e., suddenly
multiplying the wave form by $\exp (i\eta x)$. A critical tilt
(alias \textit{kick}), $\eta _{c}(P)$, beyond which the compacton
starts to move, is found. The dependence $\eta _{c}(P)$ features
variations with the same period in $P$ as the above-mentioned
switching of the quiescent solitons. Collisions between moving
compactons are also studied in Sec. III by means of direct
simulations. The paper is concluded by Sec. IV.

\section{Quasi-compactons: numerical and analytical results}

\subsection{Numerical simulations}

Modulation function $V(x)$ in Eq. (\ref{NLSP}) corresponding to Fig. \ref%
{fig_1} is a piecewise-constant one, of the Kronig-Penny type. In this
paper, we approximate $V(x)$ by the first term of its harmonic
decomposition, assuming that contributions of higher harmonics are
negligible for sufficiently broad solitons:
\begin{equation}
V(x)=\left( V_{0}/2\right) \left[ 1-\cos \left( 2\pi x{/d}\right) \right] ,
\label{potential}
\end{equation}%
where $d$ and $V_{0}>0$ are the modulation period and depth. The scaling is
set by fixing $d\equiv 20$, which leaves $V_{0}$ as a free parameter. In
this section, we present numerical results for $V_{0}=0.02,~0.03$ and $0.04$%
, which adequately illustrate the generic situation. Below, we focus
on solitons with a sufficiently large amplitude, as for small
amplitudes the truncated expansion of the saturable nonlinearity
amounts to the previously studied Kerr model \cite{Kar12,LYY1,LYY2}.

\begin{figure}[tbp]
\centering
\subfigure[]{ \label{fig_2_a}
\includegraphics[scale=0.25]{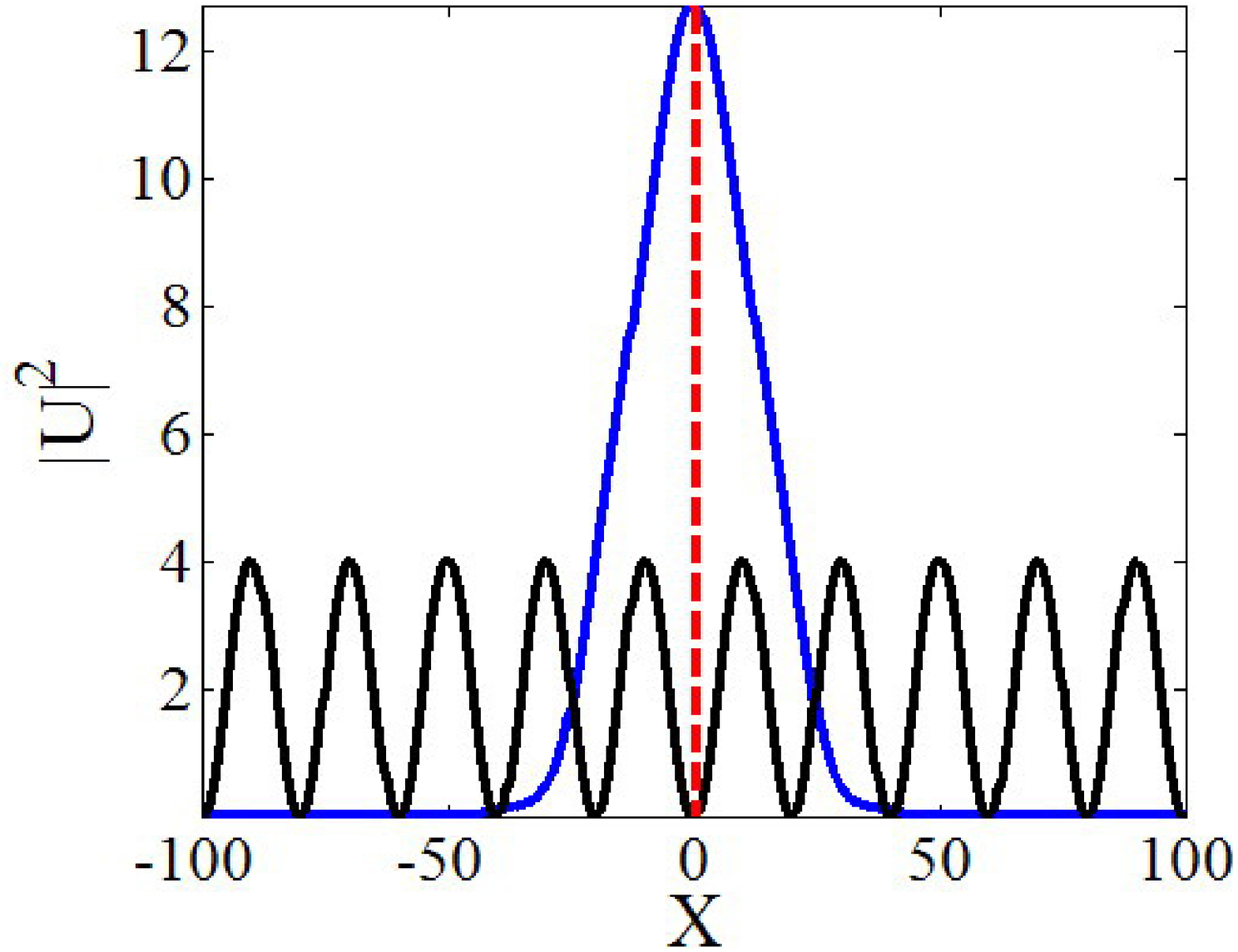}} \hspace{0.00in}%
\subfigure[]{ \label{fig_2_b}
\includegraphics[scale=0.25]{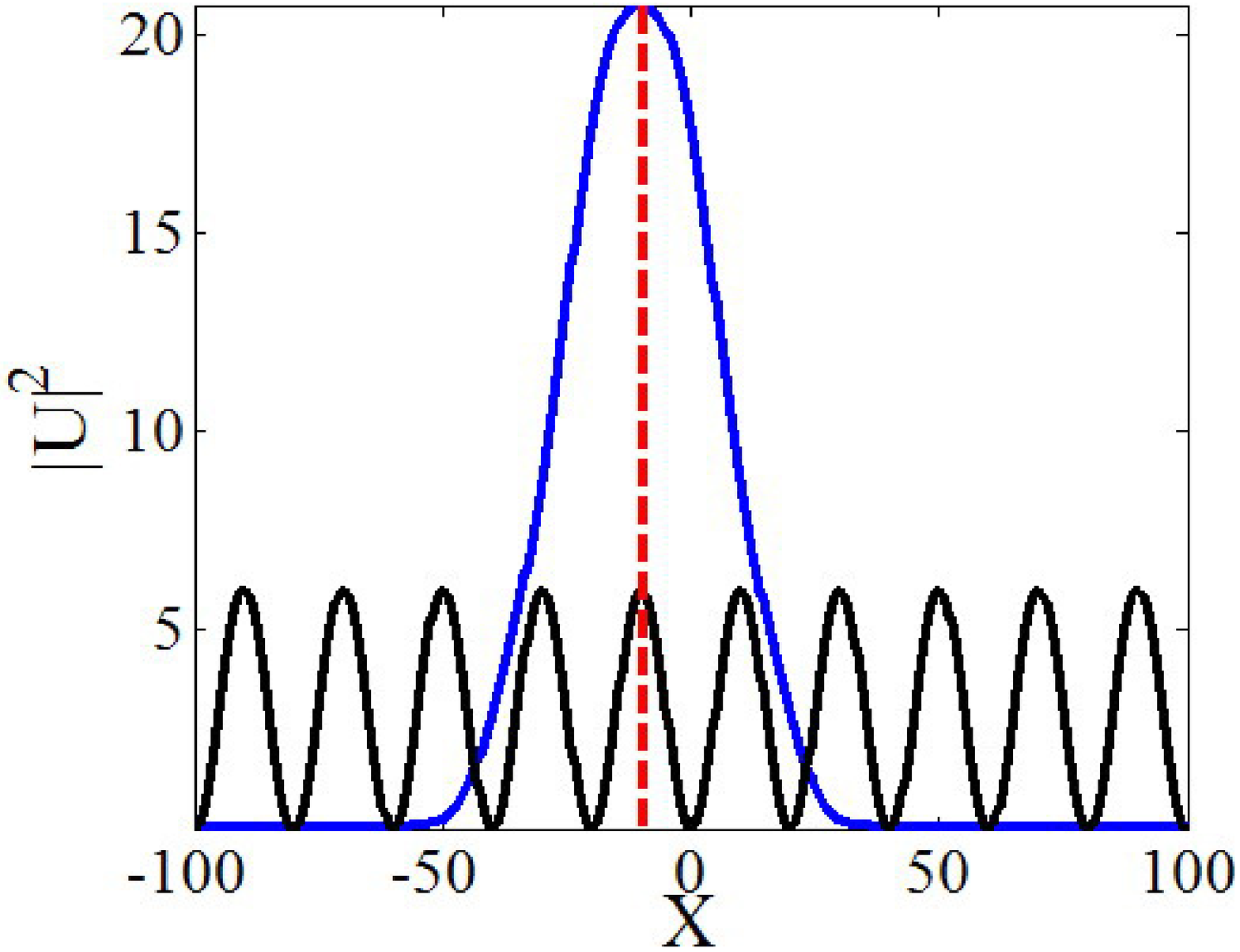}} \hspace{0.00in}
\subfigure[]{ \label{fig_2_c}
\includegraphics[scale=0.25]{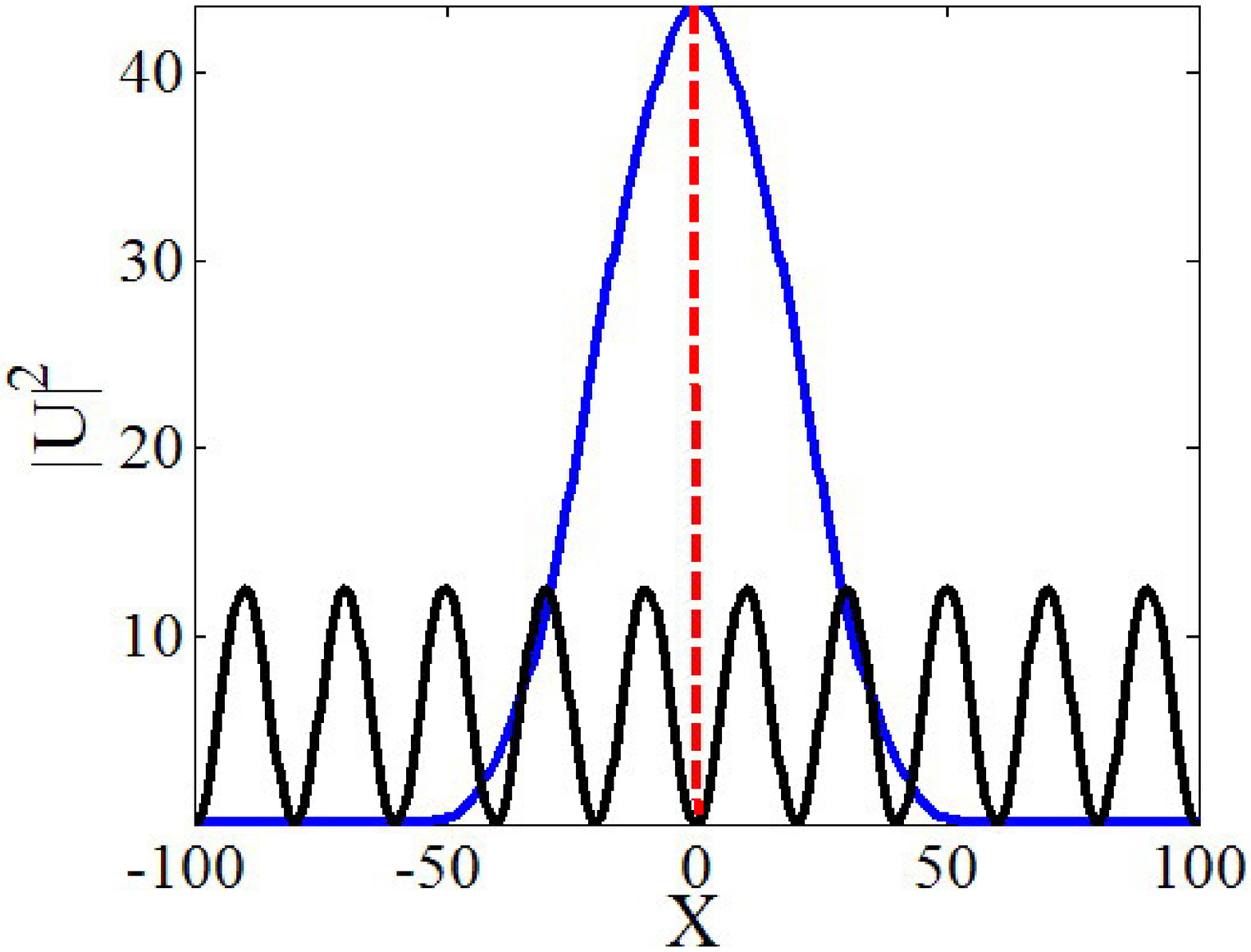}} \hspace{0.00in}
\subfigure[]{ \label{fig_2_d}
\includegraphics[scale=0.25]{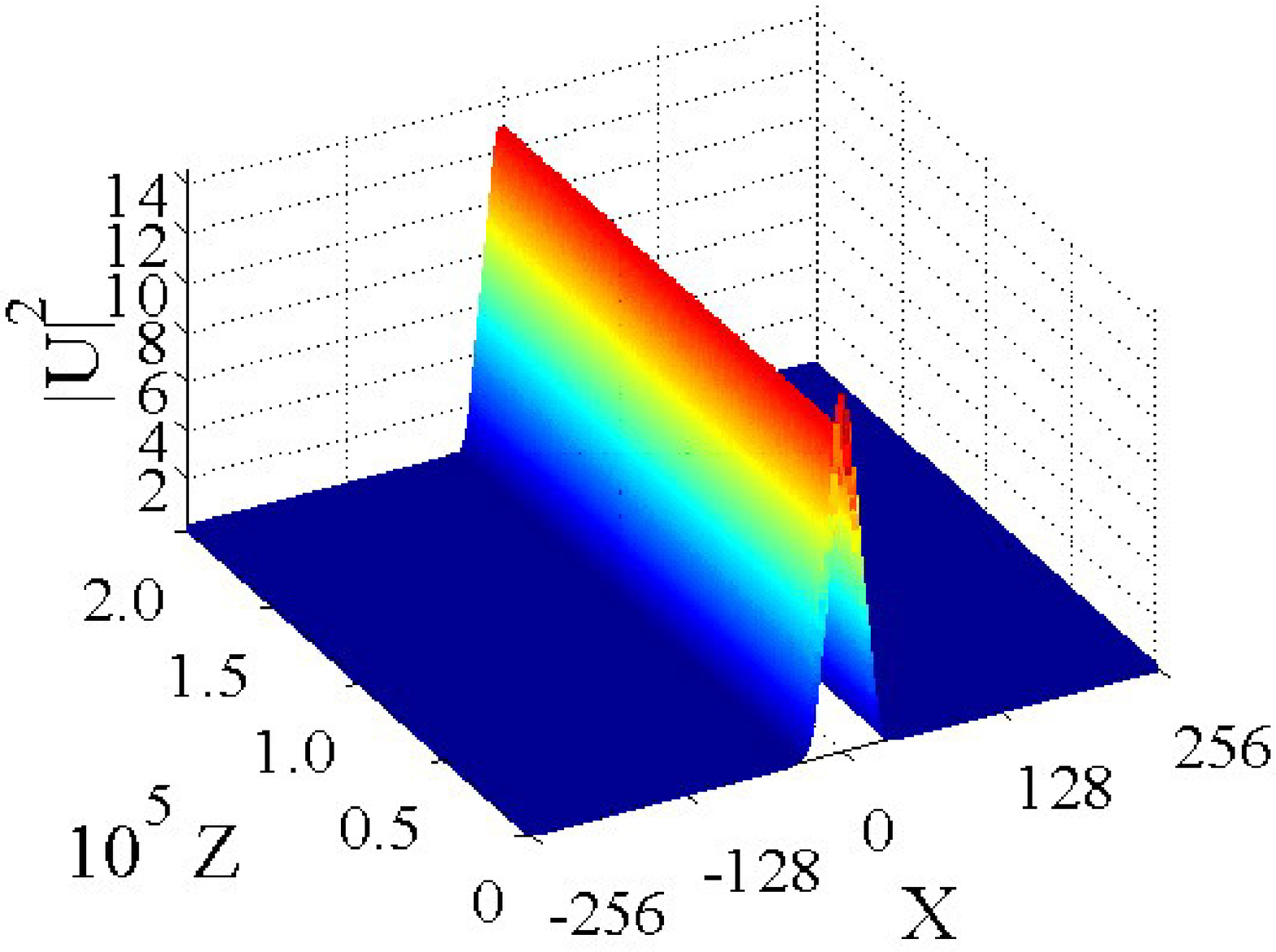}} \hspace{0.00in}
\subfigure[]{ \label{fig_2_e}
\includegraphics[scale=0.25]{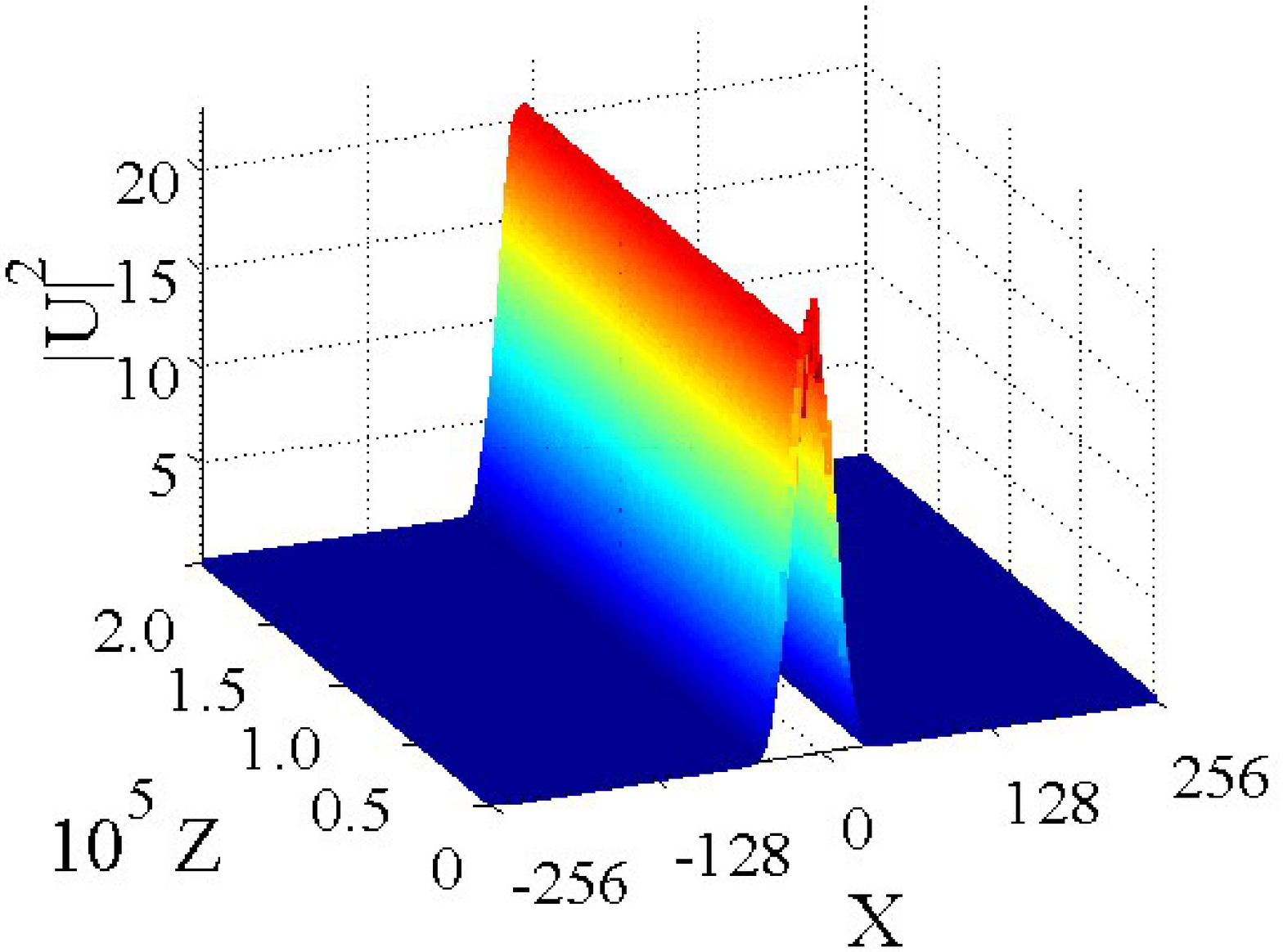}} \hspace{0.00in}
\subfigure[]{ \label{fig_2_f}
\includegraphics[scale=0.25]{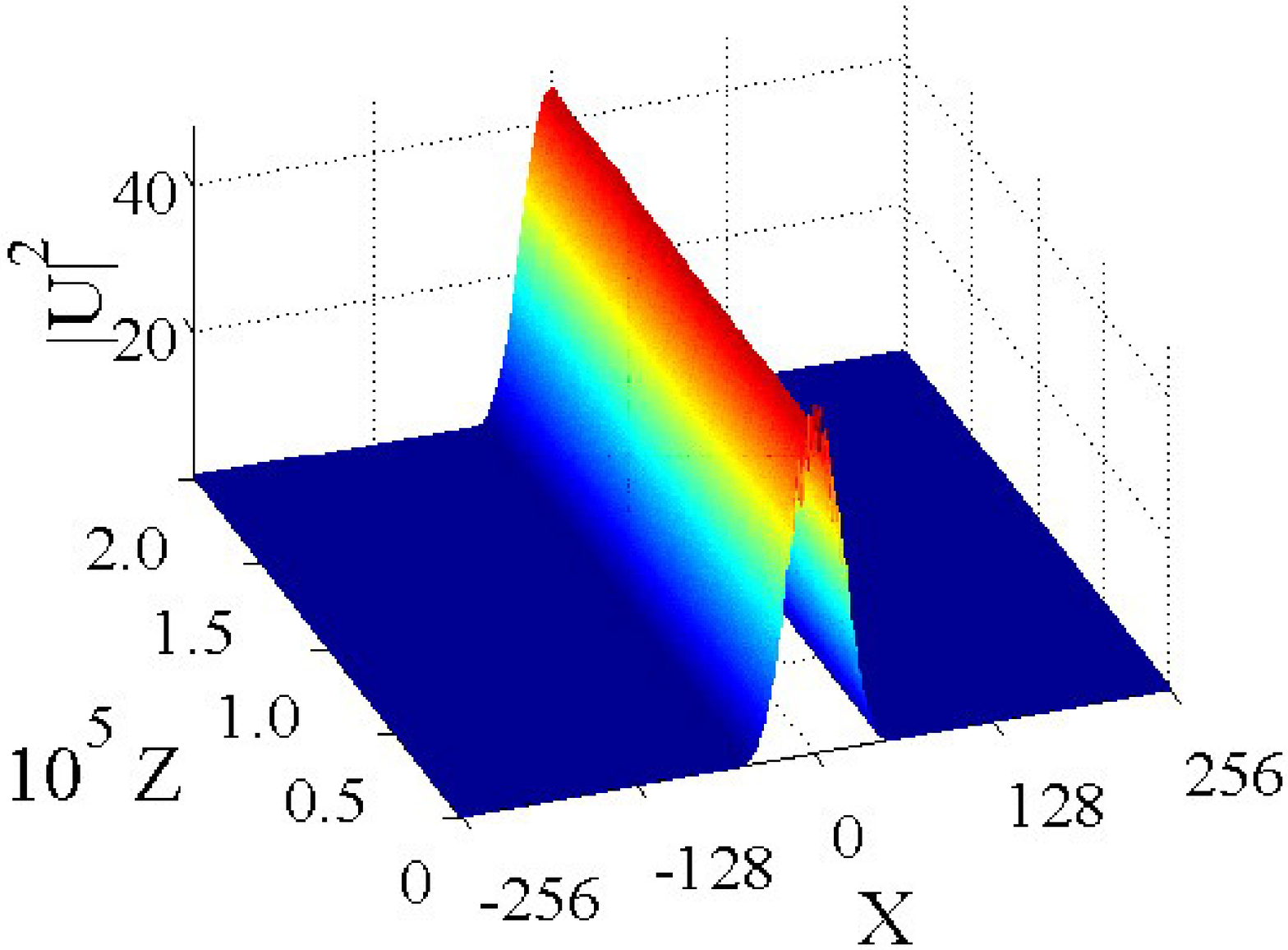}}
\caption{(Color online) Numerically found soliton profiles (blue solid
lines) for the modulation function (shown by the black solid line) with
depth $V_{0}=0.02$ (recall the modulation period is scaled to be $d\equiv 20$%
). (a) At total power $P=400$, the center of the soliton is located at the
minimum of $V(x)$. (b) At $P=800$, the soliton shifts to the maximum of $%
V(x).$ (c) At $P=2000$, the solitons shifts back to the minimum of $V(x)$.
Panels (d), (e) and (f) illustrate the stability of the solitons via direct
simulations of their perturbed evolution.}
\label{fig_2}
\end{figure}

Typical numerical results for solitons solution at different values of the
total power, $P=\int_{-\infty }^{+\infty }|u(x)|^{2}dx$, are displayed in
Fig. \ref{fig_2}. The stationary profiles, presented in Figs. \ref{fig_2_a}-%
\ref{fig_2_c}, were generated by dint of the the imaginary-time-propagation
method \cite{Chiofalo}. It is concluded that the profiles are strongly
localized, making the solitons QCs (quasi-compactons) [a solution fully
localized within a finite (\emph{compact}) interval of $y$ is usually called
a compacton \cite{Rosenau,JHH}]. On the other hand, the soliton broadens
with the increase of $P$. At relatively small values of the total power,
e.g., $P=400$, the center of the soliton is located at a minimum of $V(x)$
[Fig. \ref{fig_2_a}]; then, at $P=800$ [Fig. \ref{fig_2_b}], the position of
the soliton is switched to a maximum of $V(x)$, and at $P=2000$ it switches
back to the minimum [Fig. \ref{fig_2_c}].

To quantify power-dependent properties of the solitons, we define the
center-of-mass coordinate and average width of the soliton as follows:
\begin{eqnarray}
&&X_{\mathrm{mc}}(P)={P}^{-1}\int_{-\infty }^{+\infty }x|U(x,P)|^{2}dx
\notag \\
&&W_{\mathrm{a}}(P)={P}^{-1}\int_{-\infty }^{+\infty
}(x-x_{mc})^{2}|U(x,P)|^{2}dx,  \label{twochar}
\end{eqnarray}%
where $U(x,P)$ is the soliton solution with power $P$. Using the integration
in imaginary time, the data were collected for the solitons from $P=100$ to $%
10000$, with modulation depths $V_{0}=0.02,~0.03$ and $0.04$. Real-time
simulations of the evolution of these solutions in the framework of Eq. (\ref%
{NLSP}) demonstrate that they all are stable against perturbations.

\begin{figure}[tbp]
\centering
\subfigure[]{ \label{fig_3_a}
\includegraphics[scale=0.175]{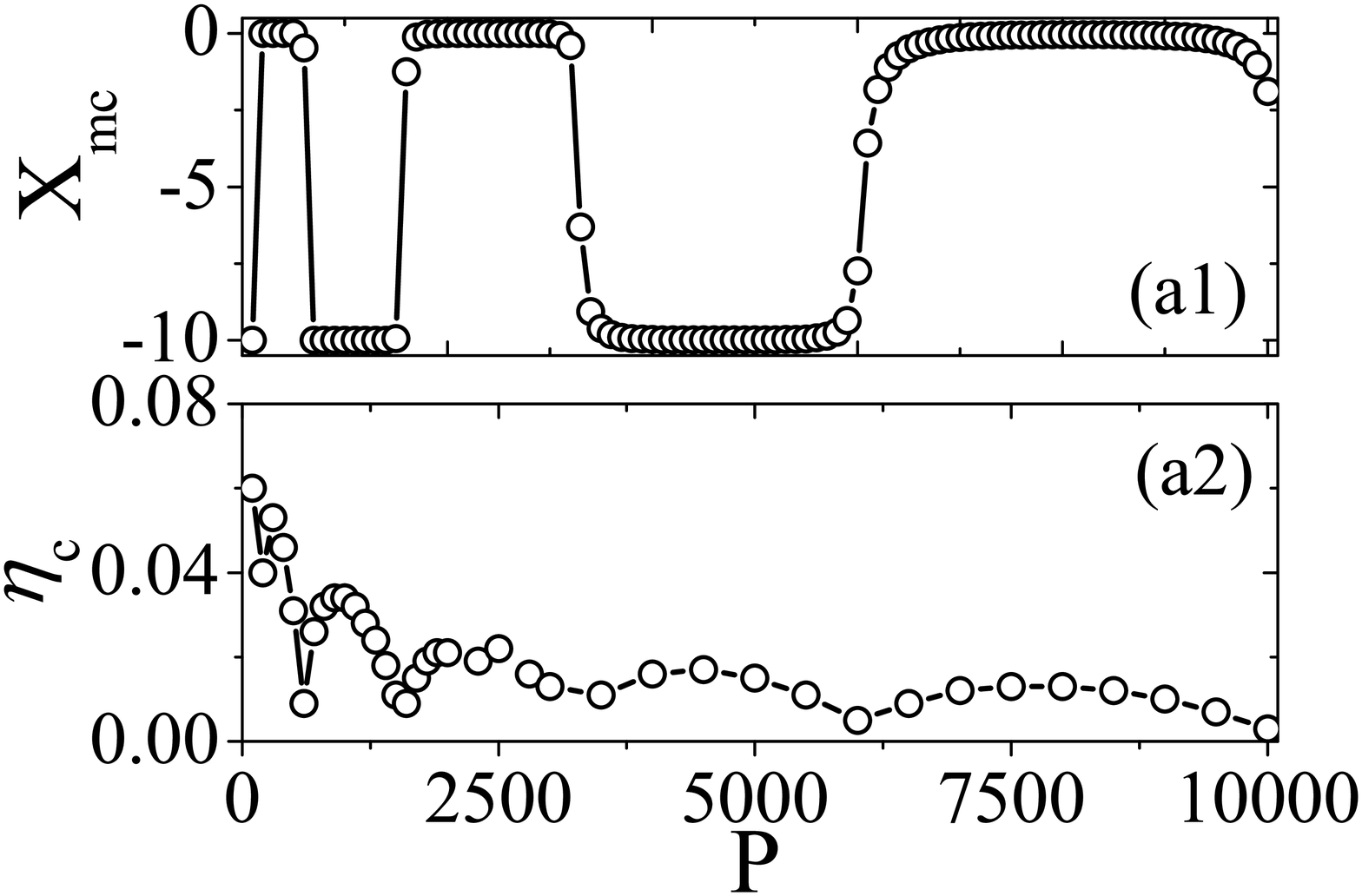}} \hspace{0.00in}%
\subfigure[]{ \label{fig_3_b}
\includegraphics[scale=0.175]{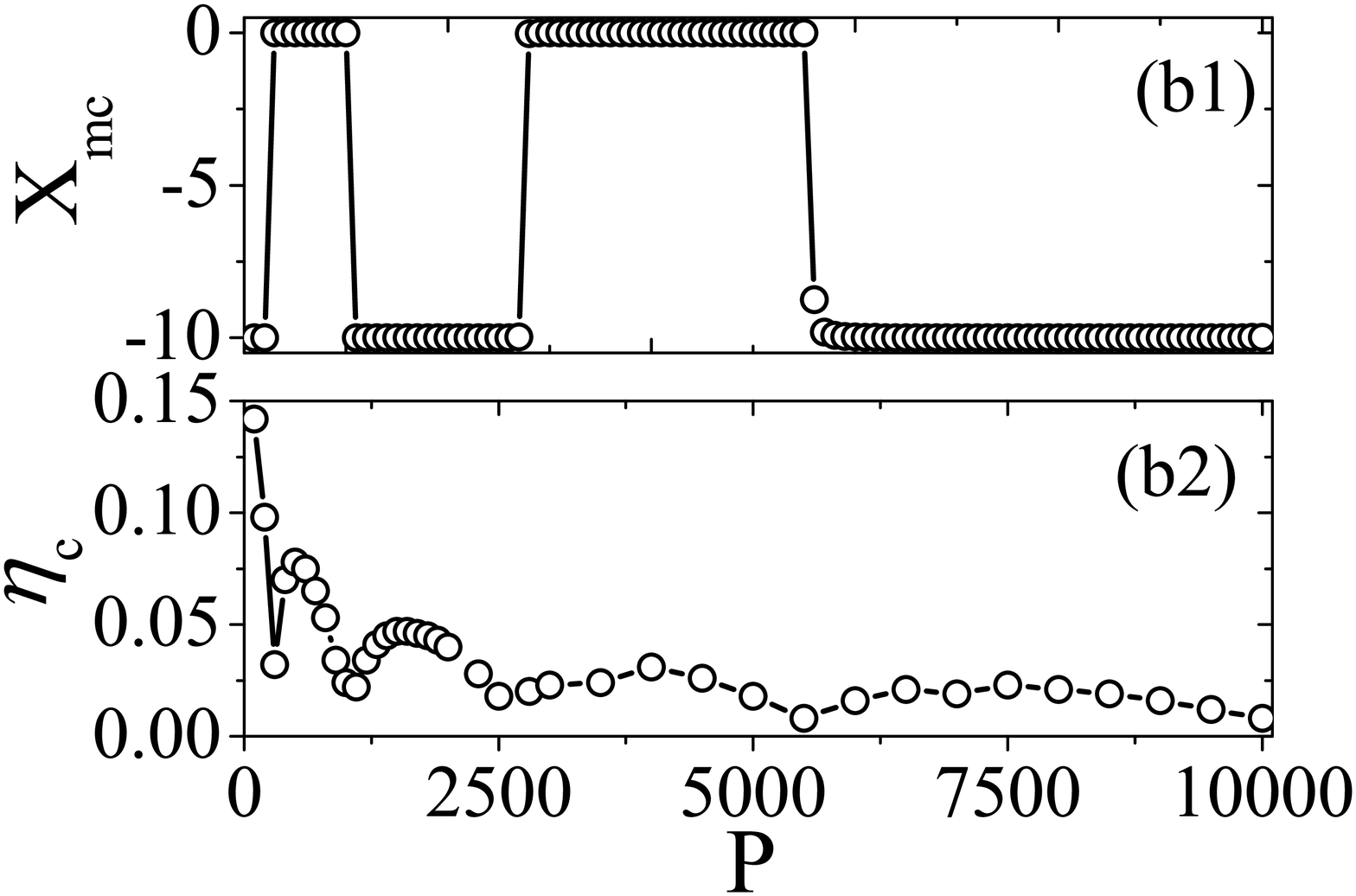}} \hspace{0.00in}%
\subfigure[]{ \label{fig_3_c}
\includegraphics[scale=0.175]{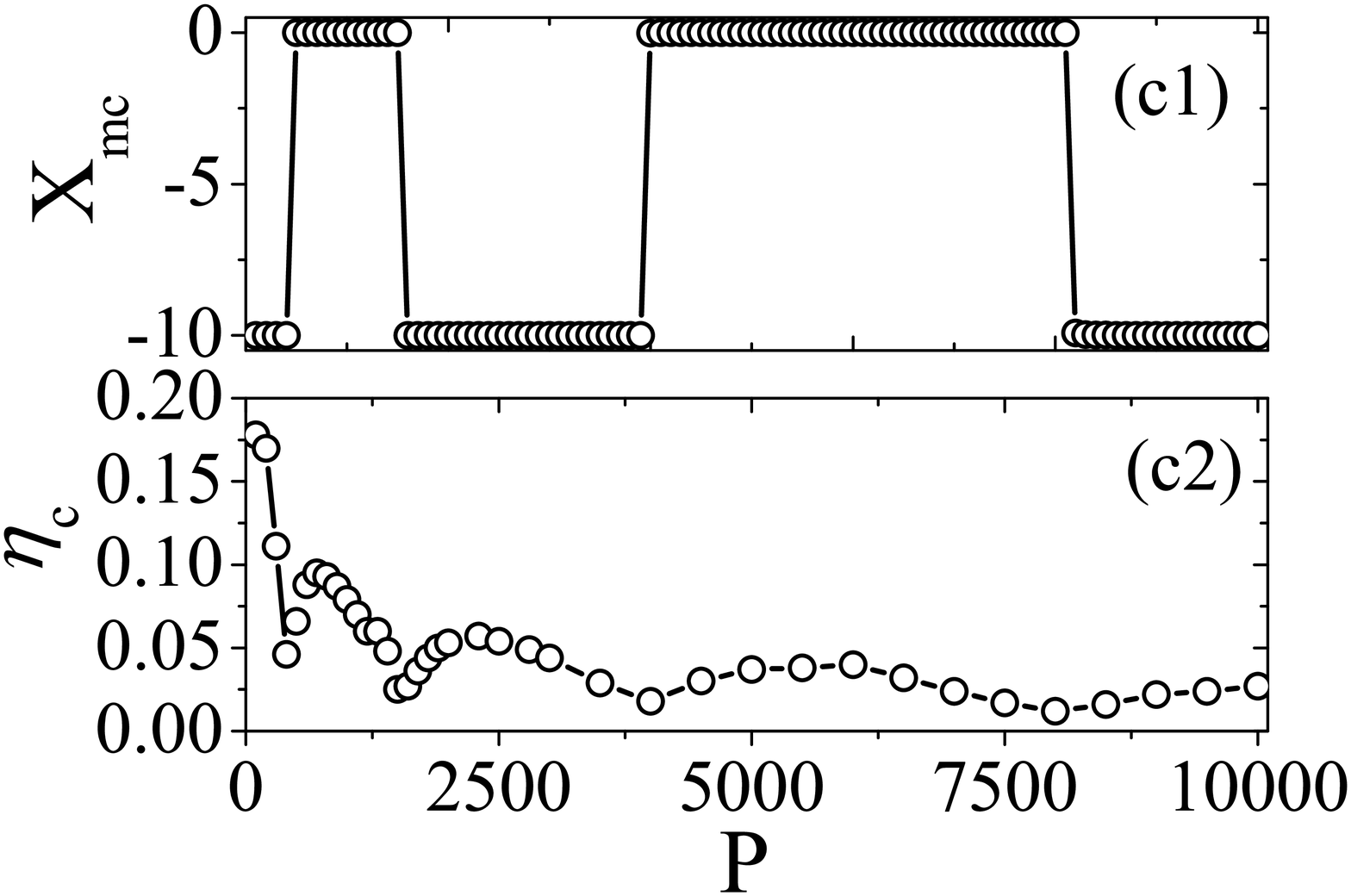}} \hspace{0.00in}%
\subfigure[]{ \label{fig_3_d}
\includegraphics[scale=0.19]{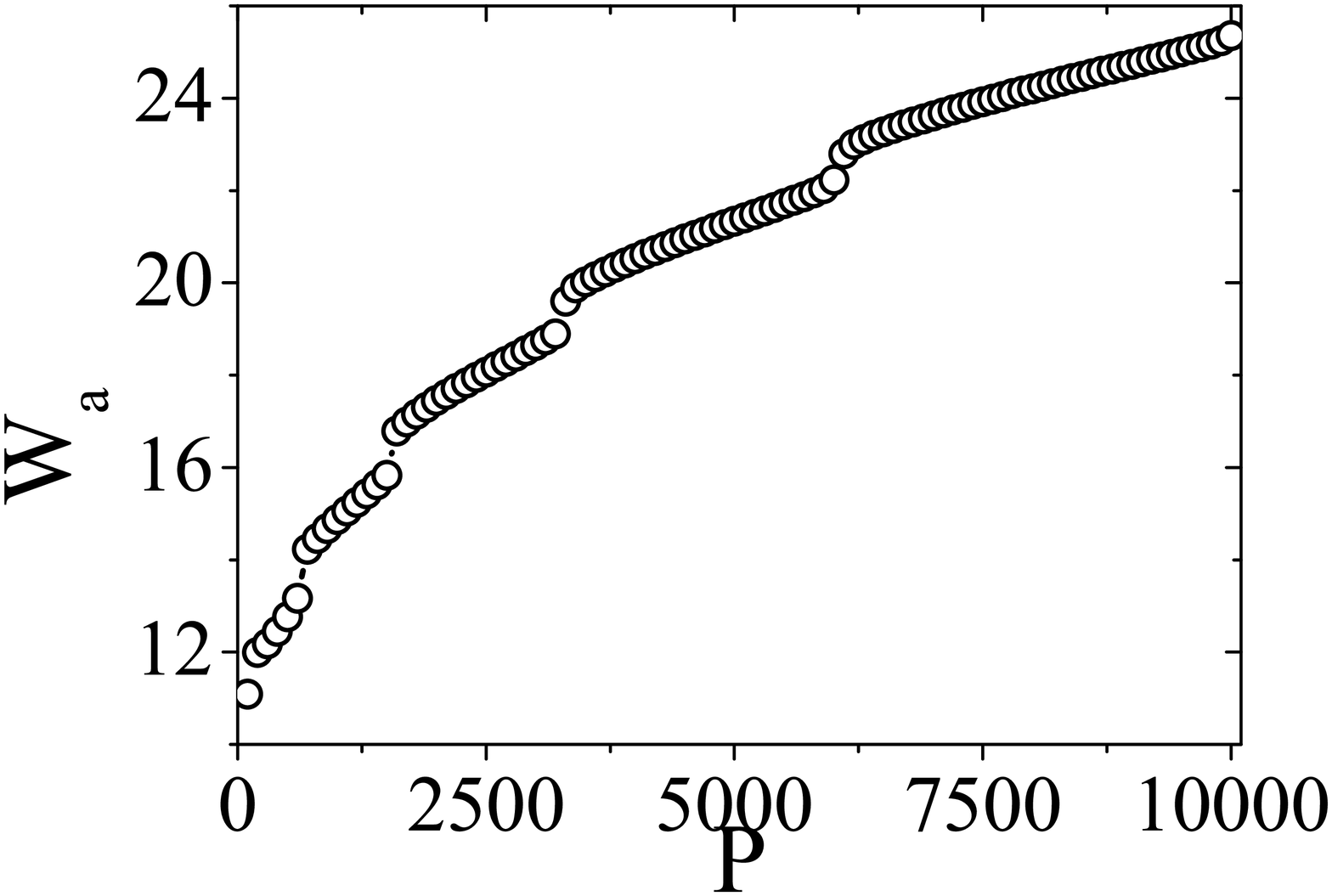}} \hspace{0.00in}%
\subfigure[]{ \label{fig_3_e}
\includegraphics[scale=0.19]{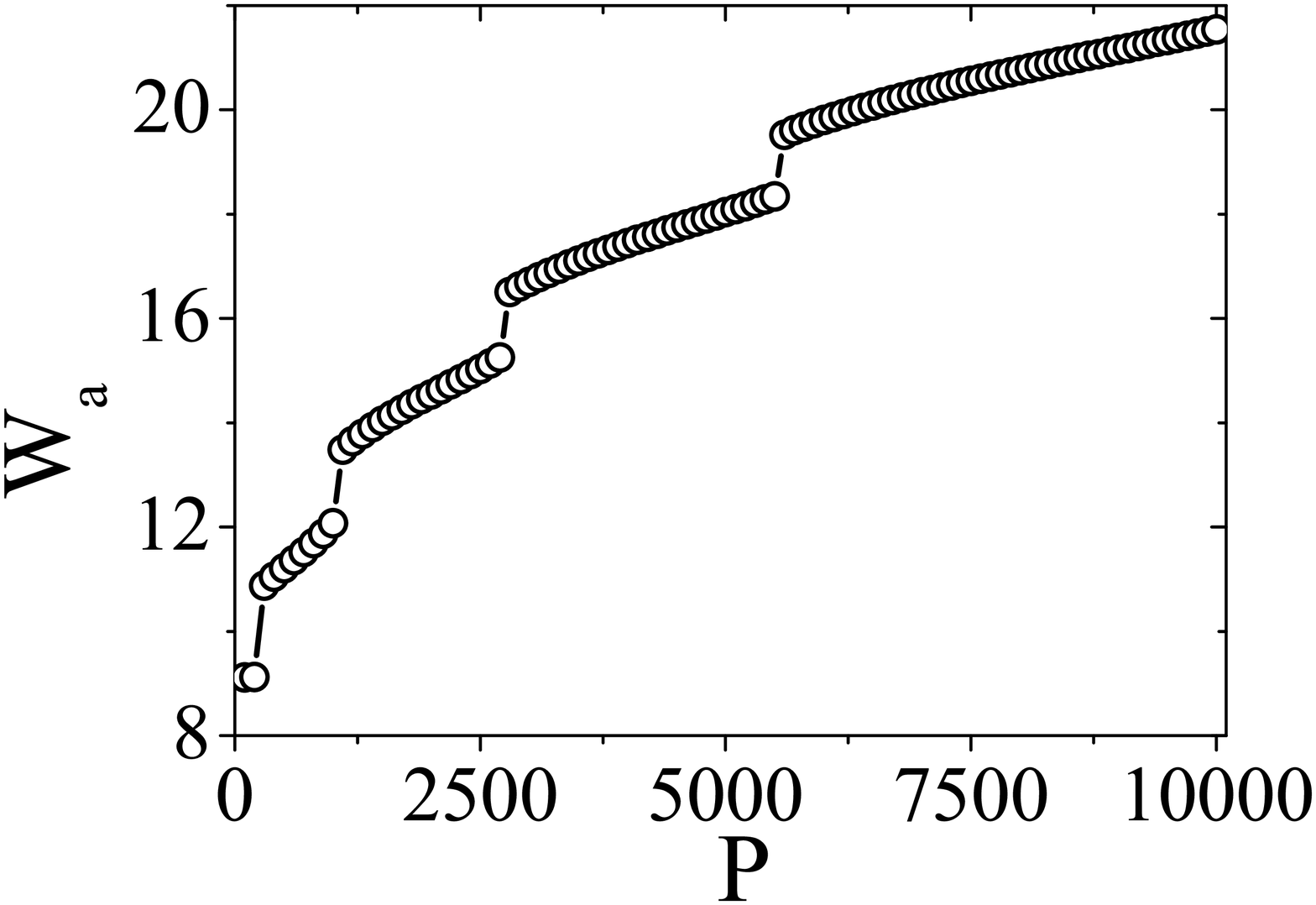}} \hspace{0.00in}%
\subfigure[]{ \label{fig_3_f}
\includegraphics[scale=0.19]{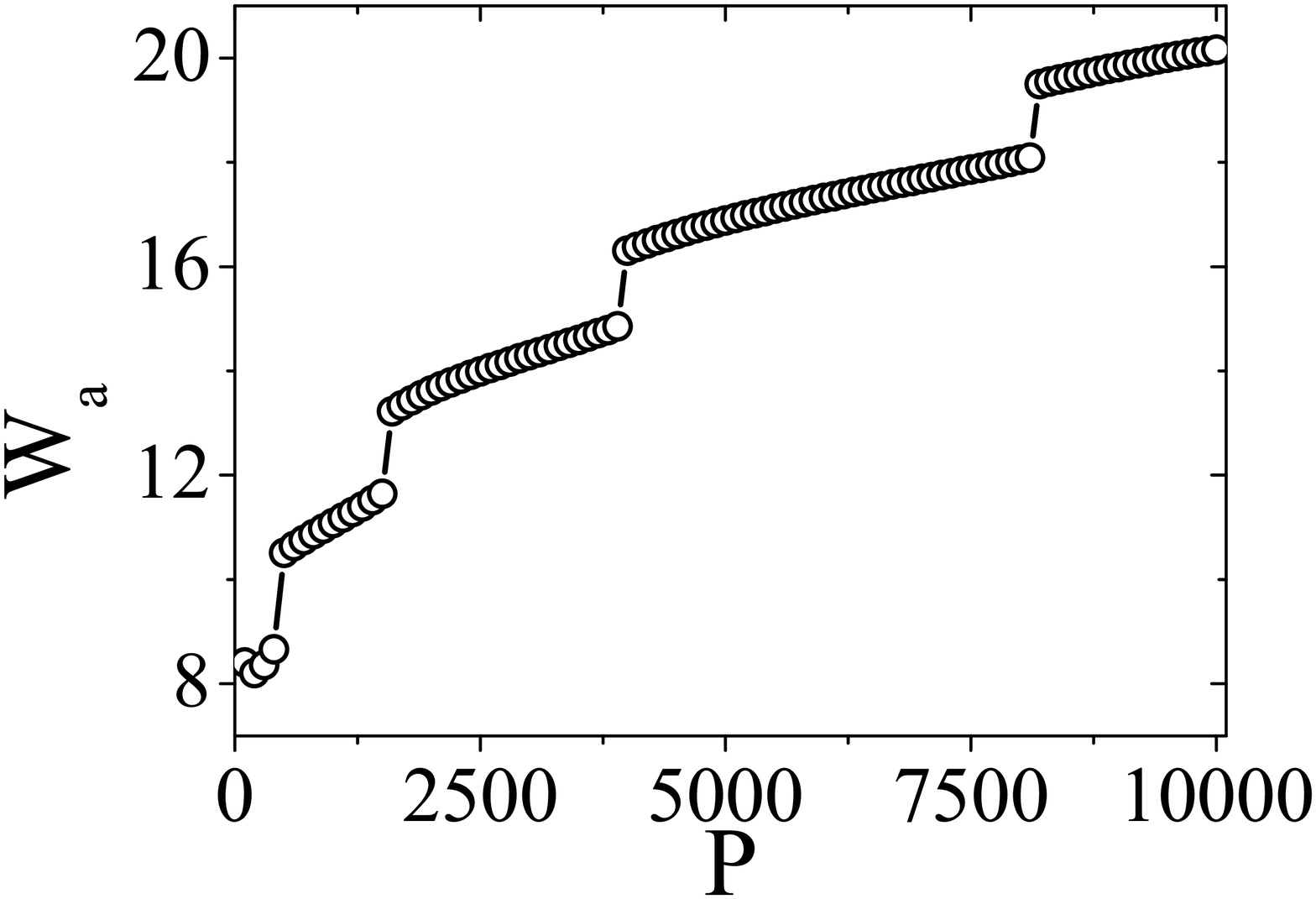}}
\caption{(Color online) The plots in panels (a1)-(c1) display the coordinate
of the soliton's center of mass, $X_{\mathrm{mc}}$, as a function of $P$,
for $V_{0}=0.02,~0.03$ and $0.04$, respectively. (a2)-(c2). The critical
kick which initiates the motion of the soliton, $\protect\eta _{\mathrm{c}}$%
, as a function of $P$ for $V_{0}=0.02,~0.03$ and $0.04$. (d)-(f) The
average width of the soliton as a function of $P$ for $V_{0}=0.02,0.03$ and $%
0.04$. }
\label{fig_3}
\end{figure}

Numerically found dependences (\ref{twochar}) are displayed in Fig. \ref%
{fig_3}. The upper panels, Fig. \ref{fig_3}(a-c), show that the soliton's
center of mass switches (as said above) between the minimum ($x=0$) and
maximum ($x=-10$) of the modulation function $V(x)$. The period of the
switching, $\Delta P$, increases with the increase of modulation depth $V_{0}
$. Further, Figs. \ref{fig_3_d}-\ref{fig_2_f} show that the average width of
the soliton grows with the increase of $P$ and decrease of $V_{0}$.

\subsection{Analytical approximations}

In saturable nonlinear media, large-amplitude solitons are always broad.
Accordingly, to develop an analytical approximation, we assume a soliton
with the width much larger than $d$. Then, in the zero-order approximation,
Eq. (\ref{NLSP}) is replaced by the equation with the averaged potential, $%
V(x)\rightarrow V_{0}/2$,
\begin{equation}
iu_{z}=-\frac{1}{2}u_{xx}+\frac{V_{0}}{2\left( 1+|u|^{2}\right) }u.
\label{V0}
\end{equation}%
Equation (\ref{V0}) gives rise to soliton solutions, $u(z,x)=e^{ikz}U(x-\xi )
$, with the coordinate of the soliton's center $\xi $, and real function $%
U(y)$ obeying the following equation:
\begin{equation}
\frac{d^{2}U}{dy^{2}}=2kU+\frac{V_{0}U}{1+U^{2}}\equiv -\frac{dW}{dU},
\label{Newton}
\end{equation}%
where $y\equiv x-\xi $. The effective potential in Eq. (\ref{Newton}),
\begin{equation}
W(U)=\allowbreak kU^{2}+\frac{V_{0}}{2}\ln \left( \frac{1+U_{0}^{2}}{1+U^{2}}%
\right) ,  \label{W}
\end{equation}%
formally corresponds to the Newton's equation of motion for a particle with
coordinate $U(y)$ ($y$ plays the role of time) and unitary mass. Then, the
shape of the soliton is determined by the respective energy equation,
\begin{equation}
\left( \frac{dU}{dy}\right) ^{2}+2k\left( U_{0}^{2}-U^{2}\right) +V_{0}\ln
\left( \frac{1+U_{0}^{2}}{1+U^{2}}\right) =0,  \label{energy}
\end{equation}%
where $U_{0}$ is the amplitude of the soliton (the largest value of $U$). As
usual, the soliton trajectory corresponds to the solution of Eq. (\ref%
{energy}) starting from $U=0$ at $y=-\infty $, bouncing back from the
potential well at $U=U_{0}$, $y=0$, and returning to $U=0$ at $y\rightarrow
+\infty $. Setting in Eq. (\ref{energy}) $U=dU/dy=0$, one can find a
relation between $k$ and $U_{0}$. Being interested in solutions with large
amplitudes, we assume here $U_{0}^{2}\gg 1$, which yields
\begin{equation}
k\approx -\left( V_{0}/2\right) U_{0}^{-2}\ln \left( U_{0}^{2}\right) ,
\label{k-appr}
\end{equation}%
\begin{equation}
U_{\min }^{2}\approx U_{0}^{2}/\ln \left( U_{0}^{2}\right) ,
\label{min-appr}
\end{equation}%
$U_{\min }$ being the coordinate of the minimum of potential
(\ref{W}). Equation (\ref{min-appr}) implies $U_{\min }^{2}\ll
U_{0}^{2}$, i.e., the minimum of the potential is located much
closer to $U=0$ than the largest value $U_{0}$. This fact suggest a
possibility to use the following approximation for solving Eq.
(\ref{energy}): at the first stage of the approximation, we drop the
logarithmic term in Eq. (\ref{energy}) altogether, i.e., we replace
the equation by its ``primitive version",
\begin{equation}
\left( \frac{dU}{dy}\right) ^{2}+2k\left( U_{0}^{2}-U^{2}\right) =0,
\label{simple}
\end{equation}%
whose obvious solution is%
\begin{equation}
U(y)=U_{0}\cos \left( \sqrt{-2k}y\right) \approx U_{0}\cos \left( \frac{%
\sqrt{V_{0}\ln \left( U_{0}^{2}\right) }}{U_{0}}y\right) ,  \label{appr}
\end{equation}%
where Eq. (\ref{k-appr}) was used to replace $k$ by the $U_{0}$.

At the second stage of the approximation, we recall that the particle does
not perform periodic oscillations, as formally follows from the expression (%
\ref{appr}), but it starts the motion from $U=0$ at $y=-\infty $, and
returns to $U=0$ at $y\rightarrow +\infty $, as the soliton solution must
do, see above. This means that approximation (\ref{appr}) is usable in
interval
\begin{equation}
|y|~<\frac{L_{\mathrm{tot}}}{2}\equiv \frac{\pi }{2\sqrt{-2k}}=\frac{\pi
U_{0}}{2\sqrt{V_{0}\ln \left( U_{0}^{2}\right) }},  \label{Ltot}
\end{equation}%
When the moving particle approaches edges of interval (\ref{Ltot}), the
logarithmic term in Eq. (\ref{energy}), which was neglected in Eq. (\ref%
{simple}), ``suddenly" becomes important, leading to the stoppage of
the particle. Thus, the full approximation in the large-amplitude
limit, $U_{0}\gg 1$, amounts to representing the soliton as the QC\
(quasi-compacton):
\begin{equation}
U(y)\approx \left\{
\begin{array}{c}
U_{0}\cos \left( \frac{\sqrt{V_{0}\ln \left( U_{0}^{2}\right) }}{U_{0}}%
y\right) ,~\mathrm{at}~~|y|~<L_{\mathrm{tot}}/2, \\
0,~\mathrm{at}~~|y|~>L_{\mathrm{tot}}/2.%
\end{array}%
\right.   \label{compact}
\end{equation}%
where $L_{\mathrm{tot}}$, given by Eq. (\ref{Ltot}), plays the role of the
total width of the QC. Note that the large amplitude $U_{0}$ of the soliton
implies that width (\ref{Ltot}) is also large, i.e., the QC is a sharply
localized but wide localized pattern. In the present approximation, the
total power of the soliton can be calculated as follows:%
\begin{equation}
P\equiv \int_{-\infty }^{+\infty }\left\vert U(y)\right\vert ^{2}dy\approx
\frac{1}{2}L_{\mathrm{tot}}U_{0}^{2}=\frac{\pi U_{0}^{3}}{2\sqrt{V_{0}\ln
\left( U_{0}^{2}\right) }}.  \label{P}
\end{equation}%
Further, using the fact that $P$ and $U_{0}^{2}$ are large, Eq. (\ref{P})
can be approximately inverted, to give the peak power (squared amplitude) of
the QC as a function of its total power:%
\begin{equation}
U_{0}^{2}\approx \left( \frac{2P}{\pi }\sqrt{\frac{2V_{0}}{3}\left( \ln
P\right) }\right) ^{2/3}.  \label{peak}
\end{equation}%
Next,using Eq. (\ref{Ltot}), it is also possible to approximately express
the full width of the compacton in terms of the total power:%
\begin{equation}
L_{\mathrm{tot}}\approx \left( \frac{3\pi ^{2}}{V_{0}}\frac{P}{\ln P}\right)
^{1/3}.  \label{width}
\end{equation}

\begin{figure}[tbp]
\centering
\subfigure[]{ \label{fig_4_a}
\includegraphics[scale=0.24]{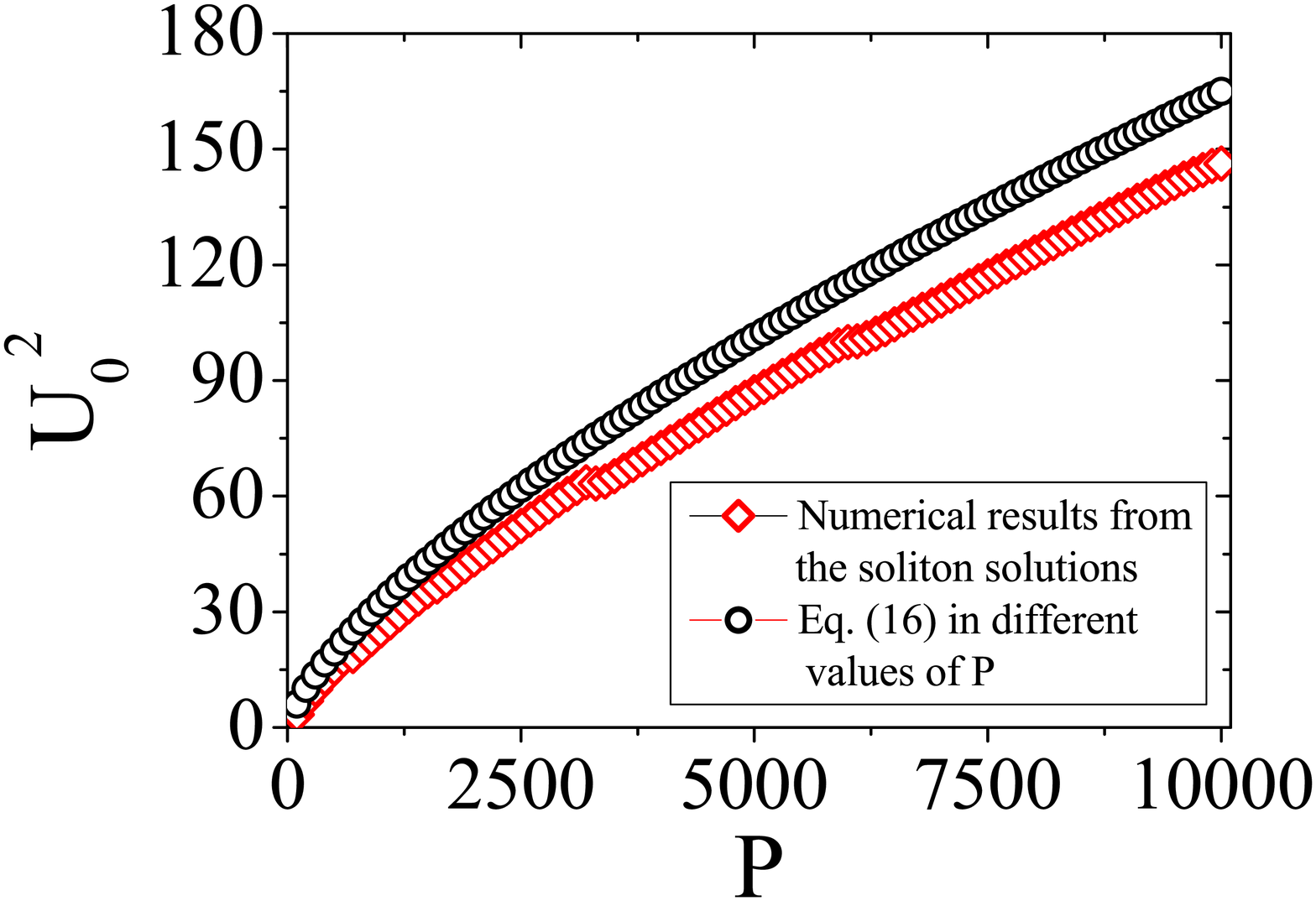}} \hspace{0.00in}%
\subfigure[]{ \label{fig_4_b}
\includegraphics[scale=0.24]{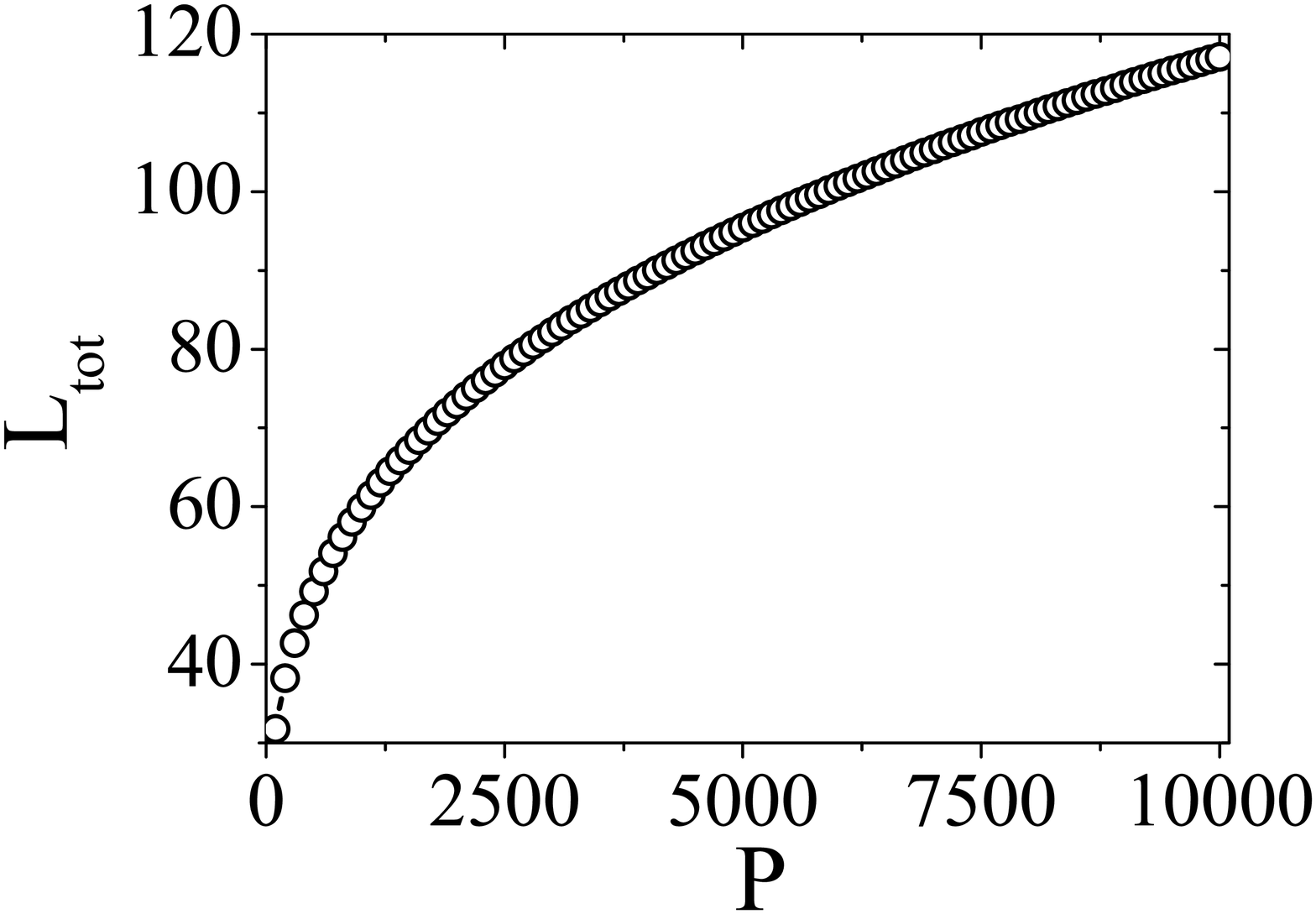}}
\caption{(Color online) (a) The comparison of the numerical results for the
peak power of the quasi-compactons with the analytical prediction given by
Eq. (\protect\ref{peak}). (b) The full width of the quasi-compacton, as
predicted by Eq. (\protect\ref{width}). In this figure, $V_{0}=0.02$.}
\label{fig_4}
\end{figure}

Relations (\ref{peak}) and (\ref{width}) are plotted in Fig.
\ref{fig_4} for different values of $P$. In particular, Fig.
\ref{fig_4}(a) demonstrates that the analytical prediction for the
peak power is consistent with the numerical
results. On the other hand, the values of the width given by Eq. (\ref{width}%
) [see Fig. \ref{fig_4}(b)] are larger than their numerically found
counterparts presented in Fig. \ref{fig_3}(d). However, one should
take into regard that width (\ref{width}) is ``all-inclusive"
(total), while the width shown in \ref{fig_3}(b) is the average one
(definitely far
smaller than the total width). Note also the similarity in the $P$%
-dependence of the width in Figs. \ref{fig_3}(d) and \ref{fig_4}(b).

To explain the position switchings of the QC, one may use the effective
energy (potential) of the interaction of the QC with the spatial modulation,
corresponding to the term $\sim \cos \left( 2\pi x/d\right) $ in expression (%
\ref{potential}). This energy can be defined as follows, with regard to the
definition of $y\equiv x-\xi $ in Eq. (\ref{Newton}):%
\begin{eqnarray}
&&E_{\mathrm{int}}(\xi )=-V_{0}\int_{-\infty }^{+\infty }dx\cos \left( \frac{%
2\pi x}{d}\right) \ln \left[ 1+U^{2}(x)\right]   \notag \\
&\approx &-V_{0}\int_{\xi -L_{\mathrm{tot}}/2}^{\xi +L_{\mathrm{tot}%
}/2}dx\cos \left( \frac{2\pi x}{d}\right) \ln \left[ 1+U^{2}(x)\right] .
\label{E}
\end{eqnarray}%
In this expression, it is taken into account that QC (\ref{compact})
occupies, approximately, a finite region of $x$, $\xi -L_{\mathrm{tot}%
}/2<x<\xi +L_{\mathrm{tot}}/2$. After three integrations by parts,
expression (\ref{E}) can be approximately calculated as follows,

\begin{equation}
E_{\mathrm{int}}(\xi )\approx K_{d}\ln (U_{0}^{2})\sin \left(
\frac{\pi ^{2}U_{0}}{d\sqrt{V_{0}\ln \left( U_{0}^{2}\right)
}}\right) \cos \left( \frac{2\pi }{d}\xi \right) ,\label{int}
\end{equation}%
where $K_{d}\equiv V_{0}^{2}d^{3}/\pi ^{3}$. Expression (\ref{int}) predicts
that the minimum of the interaction energy is located at points $\xi =nd$
with integer $n$ (at minima of the modulation function) for values of $U_{0}$
such that $\sin \left( \pi ^{2}U_{0}/\left[ d\sqrt{V_{0}\ln \left(
U_{0}^{2}\right) }\right] \right) <0$, and at points $\xi =\left(
n+1/2\right) d$ (at maxima of the modulation) in the opposite case, $\sin
\left( \pi ^{2}U_{0}/\left[ d\sqrt{V_{0}\ln \left( U_{0}^{2}\right) }\right]
\right) >0$. Because $U_{0}$ grows as the function of $P$, as per Eq. (\ref%
{peak}), this means that, with the increase of $P$, the position of the
energy minimum must indeed switch between adjacent extrema of the modulation
function. The fact that $V_{0}$ appears in the denominator of the argument
of the sinusoidal factor explains why the period of the switching increases
with $V_{0}$.

\section{Mobility of the quasi-compactons}

\subsection{Driving the soliton by the phase tilt}

Mobility of solitons under the combined action of LL and NL is a problem
which was considered in a number of different settings \cite%
{Sakaguchi,Zhou,Kar3,Oster,RMP_Kar}. To set the QC in motion, we
follow the standard approach, suddenly kicking a quiescent one,
i.e., multiplying the respective solution, $U(x)$, by the phase-tilt
factor: $U(x,\eta)\rightarrow \exp (i\eta x)U(x)$, where $\eta $ is
the strength of kick.

\begin{figure}[tbp]
\centering
\subfigure[]{ \label{fig_5_a}
\includegraphics[scale=0.3]{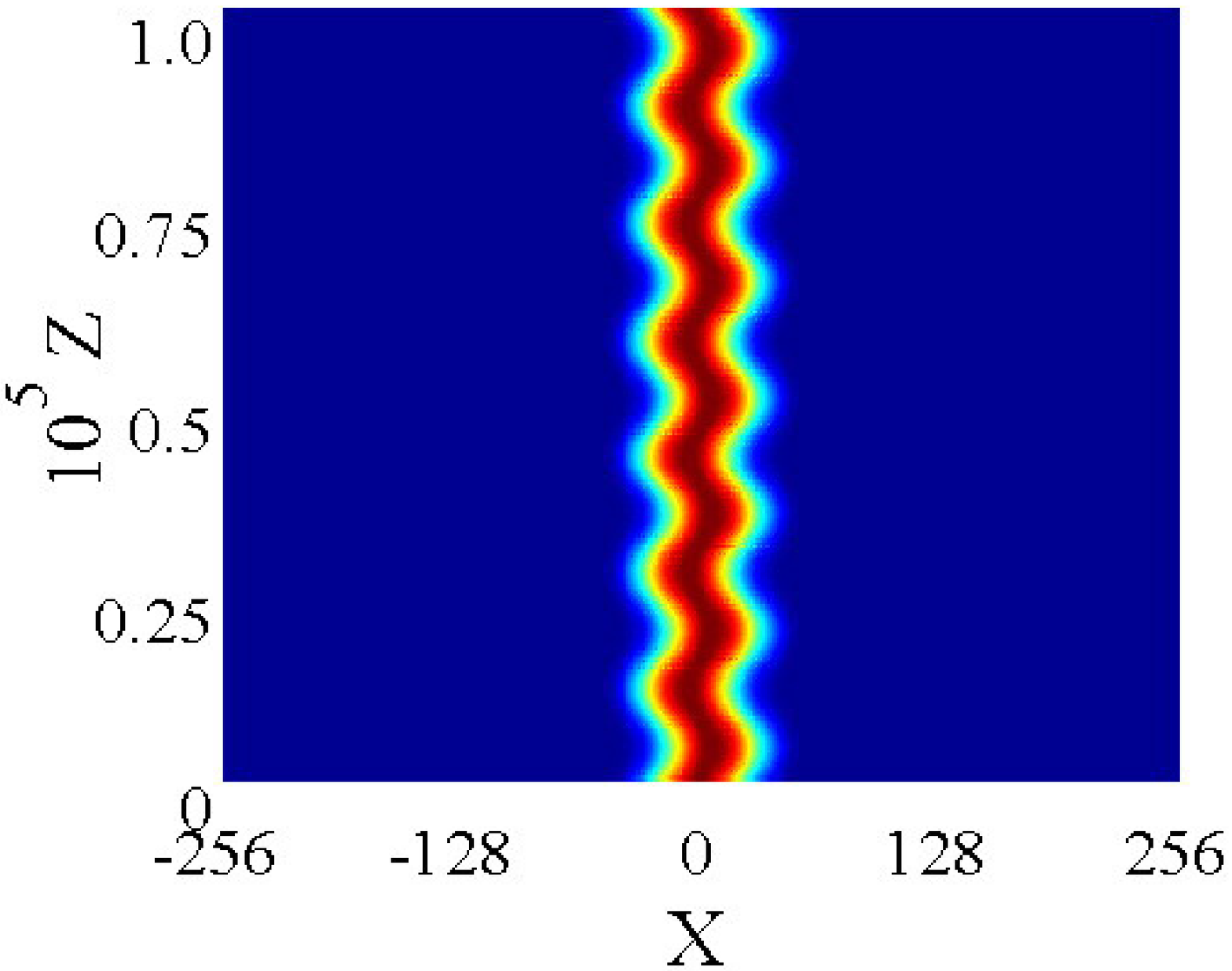}} \hspace{0.00in}%
\subfigure[]{ \label{fig_5_b}
\includegraphics[scale=0.3]{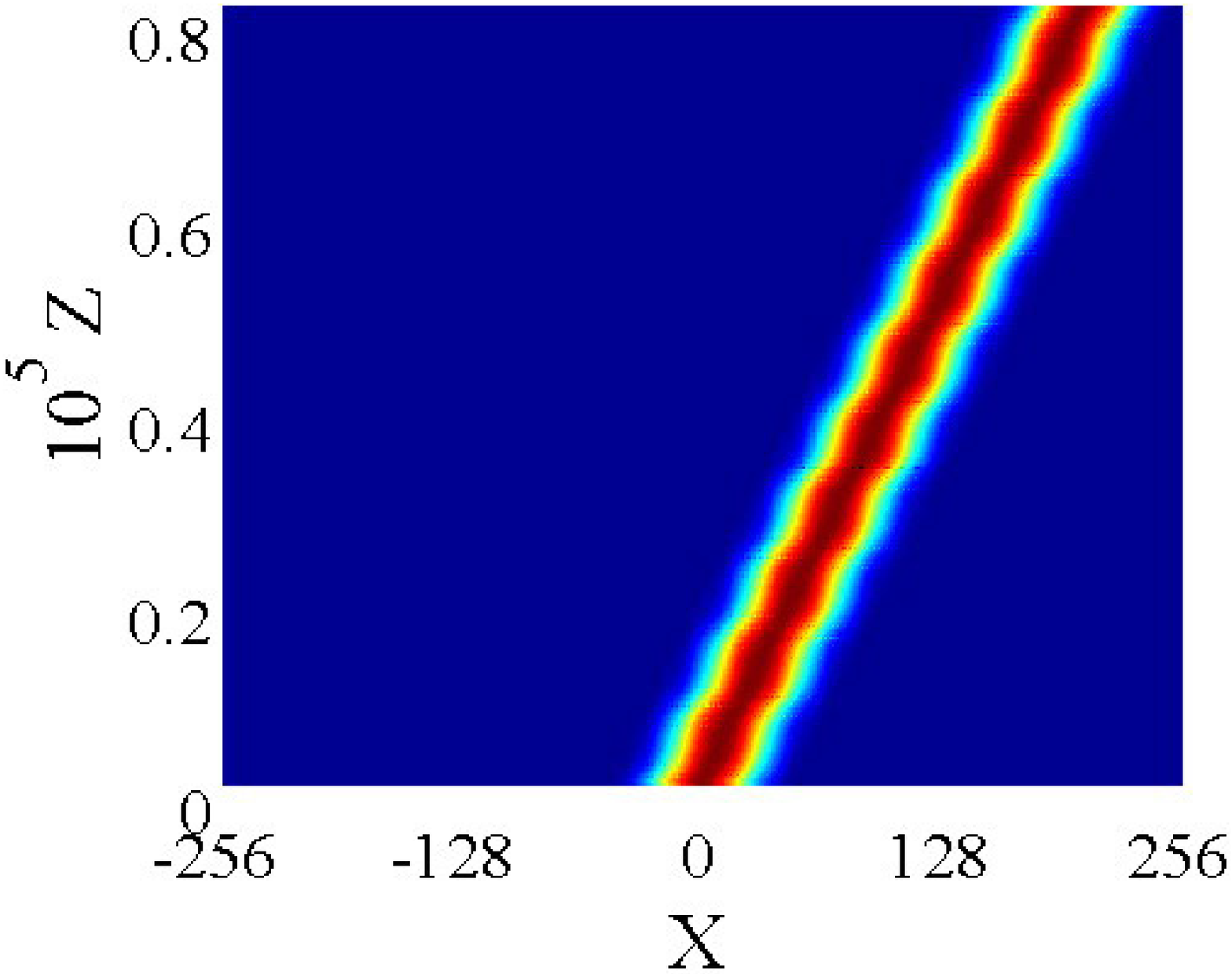}}
\caption{(Color online) Direct simulations of the compacton with
$P=2000$, initiated by the application of the phase tilt with $\eta =0.020(%
\pi /d)$ (a) and $\protect\eta =0.025(\pi /d)$ (b). The lattice
depth is $V_{0}=0.02$.} \label{fig_5}
\end{figure}

The simulated propagation of the QC, induced by kicking the standing
one, $U(x,P=2000)$, is displayed in Fig. \ref{fig_5}, for two
different values of $\eta $. It is observed that $\eta
=0.020(\pi/d)$ causes only oscillations of the soliton, without any
progressive motion. However, a slightly higher kick,
$\eta=0.025(\pi/d)$, is sufficient to set the same soliton in
persistent motion. These results imply that there must be a
threshold (critical) value of the kick, $\eta _{\mathrm{c}}$, such that the kick with $%
\eta >\eta _{\mathrm{c}}$ makes the given QC a traveling soliton. For $P=2000$%
, it is $\eta _{\mathrm{c}}=0.021\left( \pi /d\right) $.

The critical kick for the QCs is shown as a function of the total power in
Figs. \ref{fig_3}(a2)-\ref{fig_3}(c2), for $V_{0}=0.02,0.03$, and $0.04$.
The general trend to the decrease of $\eta _{\mathrm{c}}$ with the increase
of $P$ is explained by the broadening of the soliton with the increase of
the total power in the case of the saturable nonlinearity, and ensuing
enhancement of the solitons' mobility \cite{Oster}. A specific feature of
the INPhC is that oscillations of $\eta _{\mathrm{c}}$ are superimposed on
top of the gradual decay of the critical kick, the period of the
oscillations coinciding with that of the switching of the center-of-mass
position. Figures \ref{fig_3_a}-\ref{fig_3_c} also demonstrate that $\eta _{%
\mathrm{c}}$ decreases with the decreases of modulation depth $V_{0}$. The
latter trend is correlated with the fact that, as seen in Figs. \ref{fig_3_d}%
-\ref{fig_3_f}, the width of the solitons increases, making them
more mobile, with the decrease of $V_{0}$ (as well as with the
increase of $P$).

\begin{figure}[tbp]
\centering 
\includegraphics[scale=0.3]{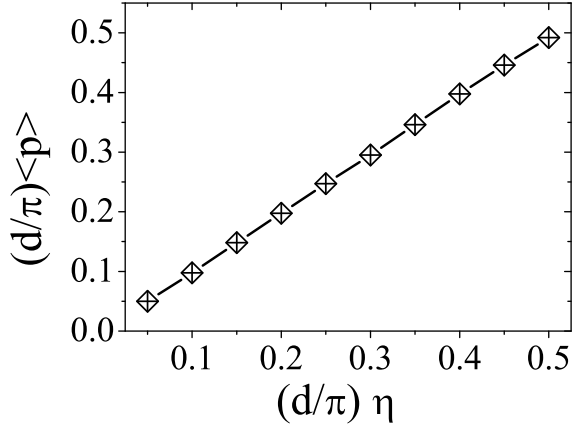}
\caption{The momentum of the soliton with $P=2000$, measured at
$z=1000$, as a function of initial kick $\eta$.} \label{fig_x}
\end{figure}

Further, Fig. \ref{fig_x} shows that the momentum of the moving
soliton, which is defined as
\begin{eqnarray}
\langle p(z)\rangle=-i\int dx
U(x,z,\eta)^{\ast}U_x(x,z,\eta),
\end{eqnarray}
is, naturally, a linear function of the initial kick, $\eta$.
Further simulations (not shown here) demonstrate that the QC are
\emph{not} destroyed even by a very strong kick, far exceeding $\eta
_{\mathrm{c}}$.

\begin{figure}[tbp]
\centering
\subfigure[]{ \label{fig_6_a}
\includegraphics[scale=0.3]{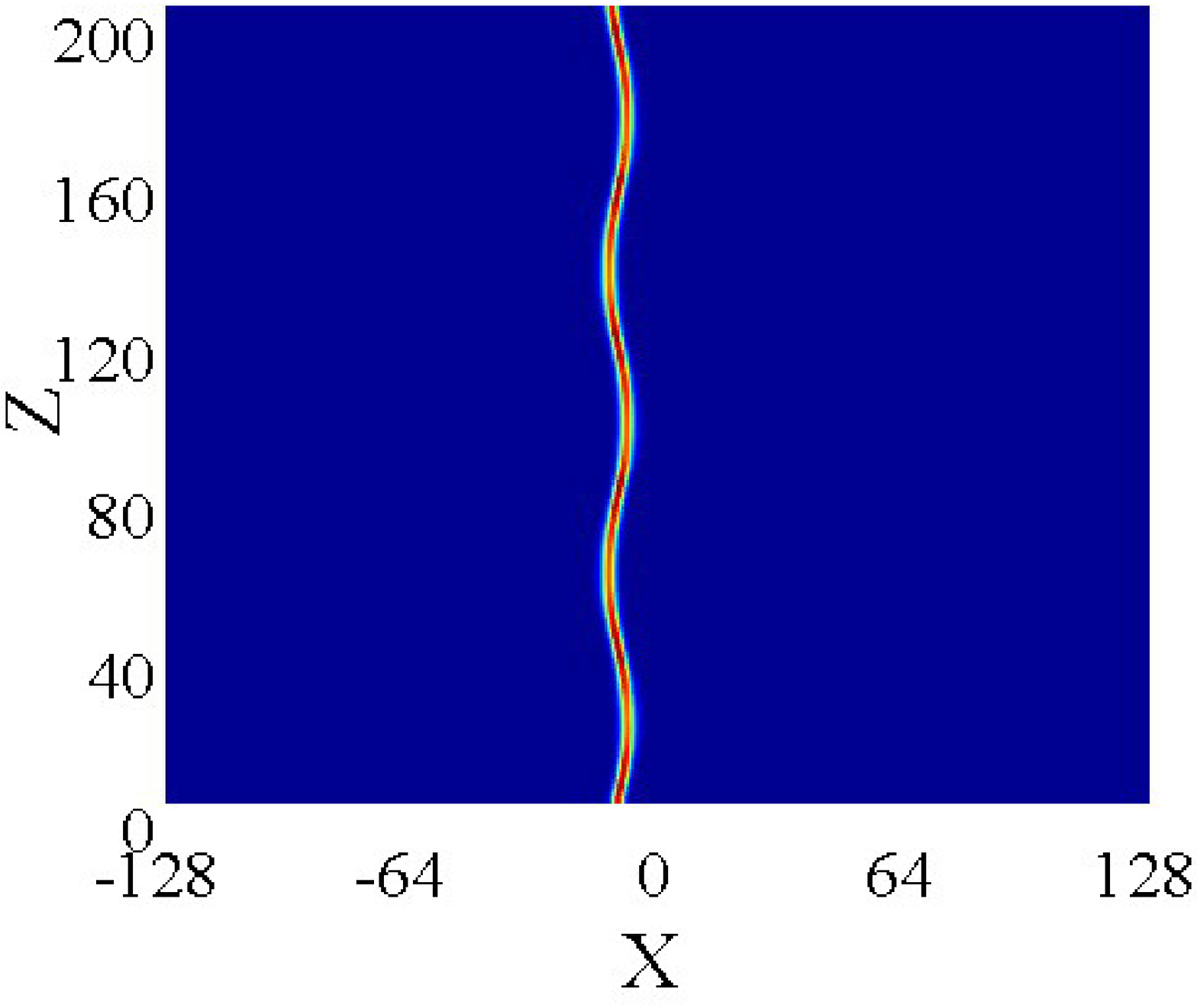}} \hspace{0.00in}%
\subfigure[]{ \label{fig_6_b}
\includegraphics[scale=0.3]{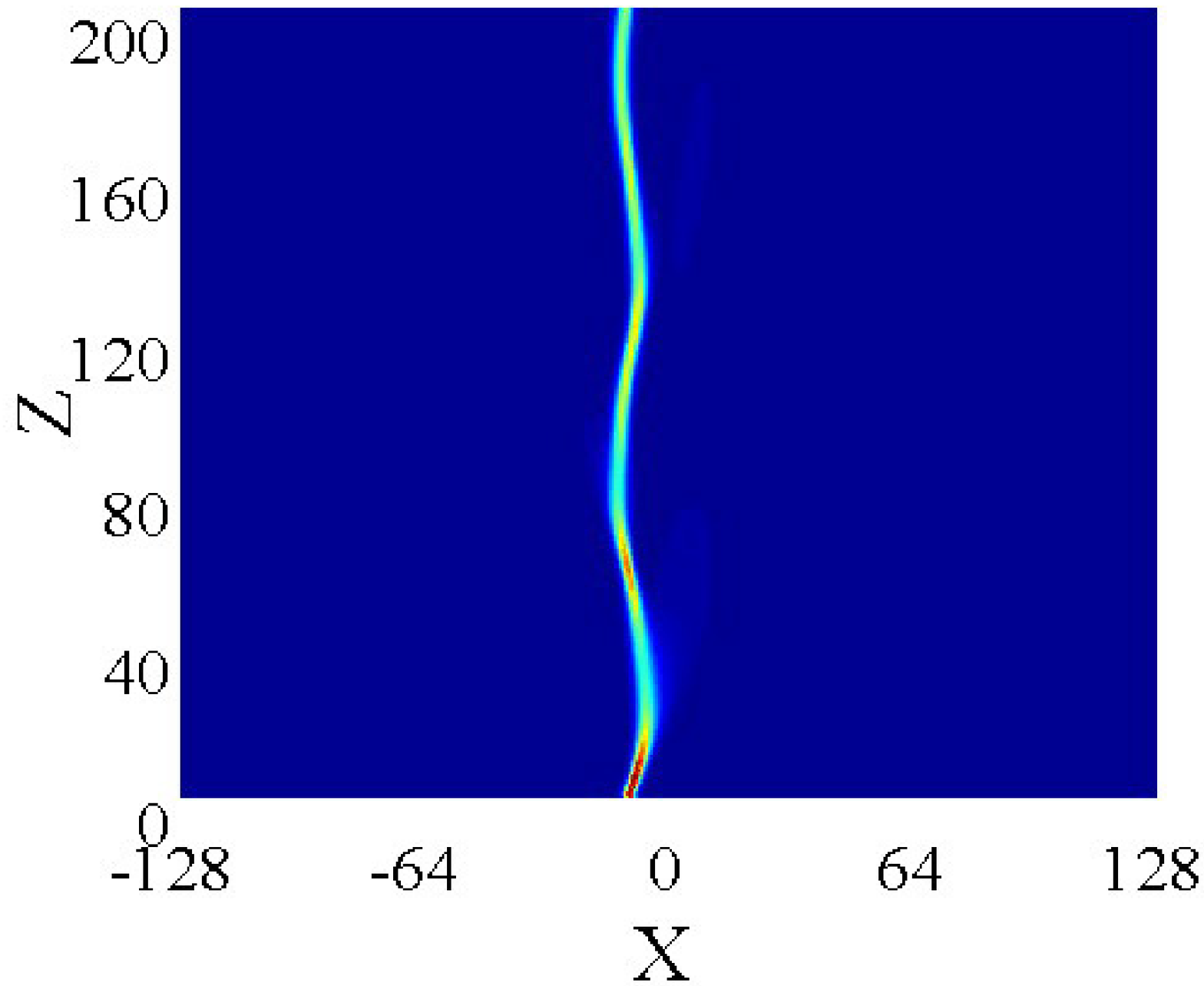}} \hspace{0.00in}%
\subfigure[]{ \label{fig_6_c}
\includegraphics[scale=0.3]{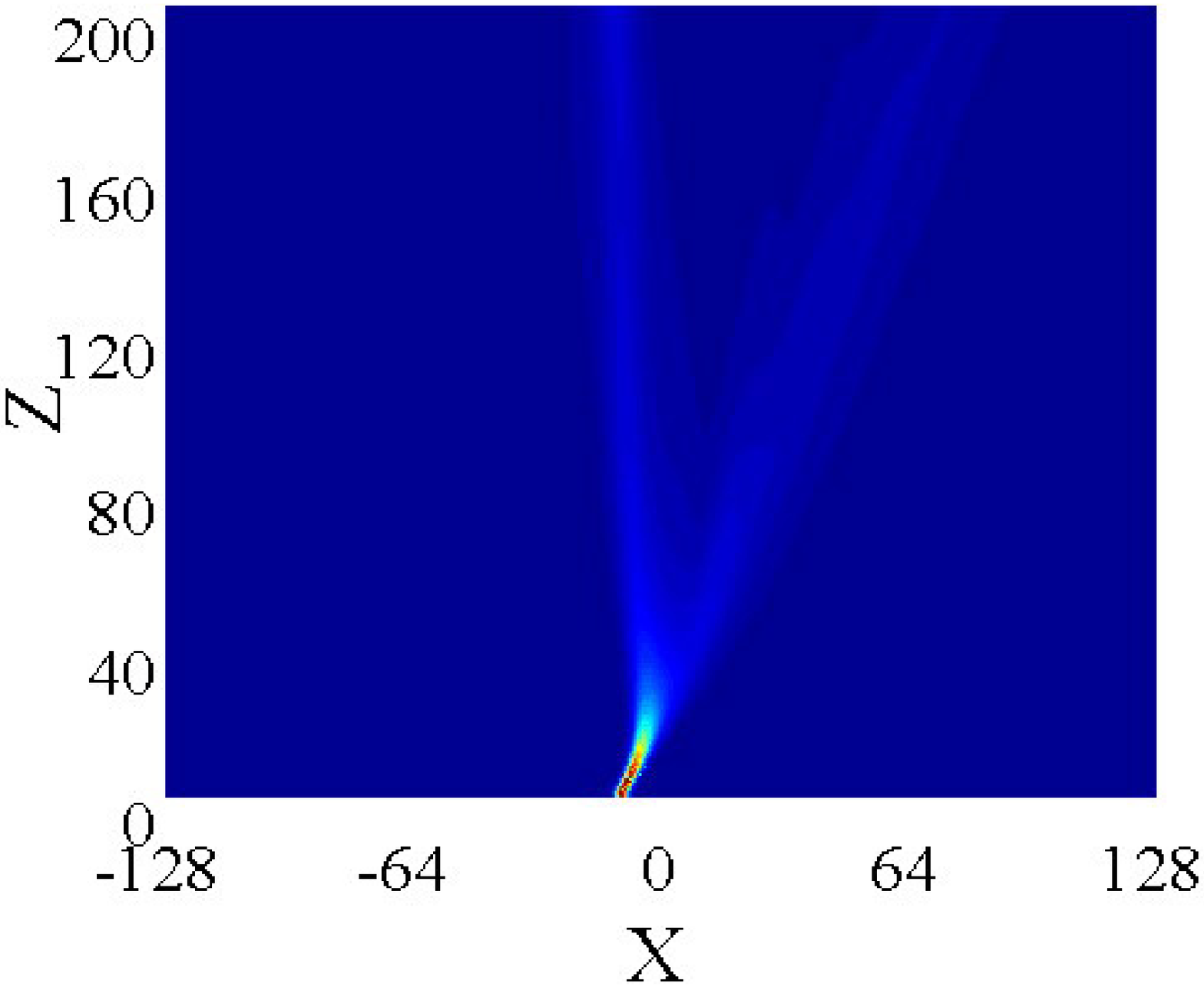}} \hspace{0.00in}%
\subfigure[]{ \label{fig_6_d}
\includegraphics[scale=0.22]{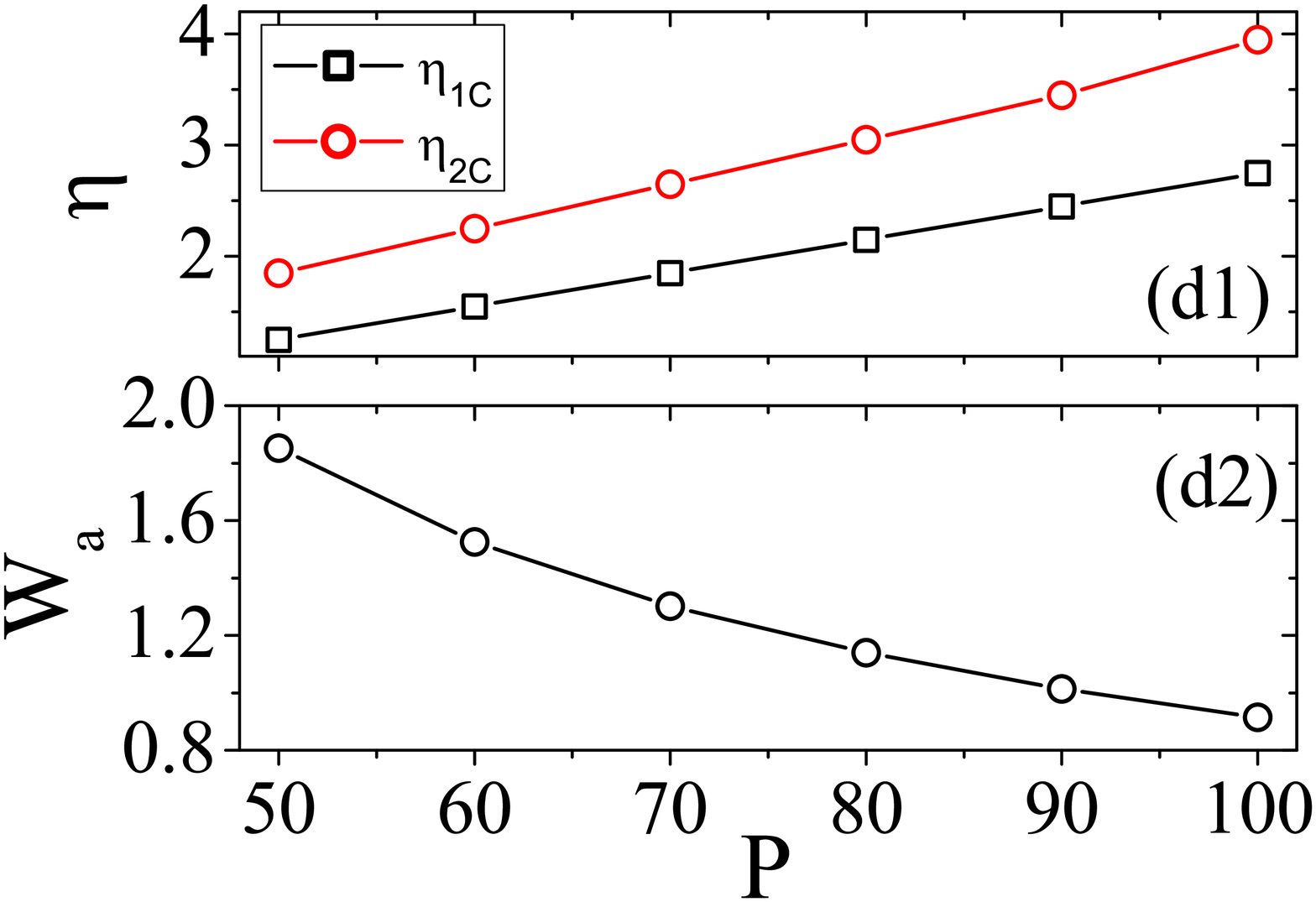}}
\caption{(Color online) (a) Direct simulations of the evolution of the
soliton with total power $P=80$ in the model with the cubic nonlinearity
[Eq. (\protect\ref{cubic})], initiated by the kick with strengths $\protect%
\eta =0.15(\protect\pi /d)$ (a), $0.25(\protect\pi /d)$ (b), and $\protect%
\eta =0.35(\protect\pi /d)$ (c). (d) The top plot: $\protect\eta _{1\mathrm{c%
}}$ is the border between stable solitons and fuzzy beams; $\protect\eta _{2%
\mathrm{c}}$ is the border between the fuzzy beams and completely destroyed
ones. The bottom plot shows the average width of the quiescent soliton
versus its total power $P$. }
\label{fig_6}
\end{figure}

It is relevant to compare the mobility of large-amplitude solitons in the
INPhCs models with the saturable and cubic self-focusing. In the latter
case, Eq. (\ref{NLSP}) is replaced by \cite{Thaw}
\begin{equation}
iu_{z}=-{\frac{1}{2}}u_{xx}+V(x)(1-|u|^{2})u,\label{cubic}
\end{equation}%
where we chose the modulation function, $V(x)$, in the same form as in Eq. (%
\ref{potential}), with $V_{0}=0.02$ and $d=20$. A typical example of
simulations of the mobility of solitons in this model is displayed in Figs. %
\ref{fig_6}(a)-(c), for three kicks, $\eta =0.15{\pi /d},0.25{\pi /d}$, and $%
0.35{\pi /d}$. In particular, Fig. \ref{fig_6_a} shows that $\eta =0.15{\pi
/d}$ give rise to oscillations of the soliton, without depinning. Further,
it is observed in Fig. \ref{fig_6_b} that the stronger kick, $\eta =0.25{\pi
/d}$, transforms the soliton into an oscillatory beam with a fuzzy shape,
but, still, it was not set in motion. Finally, Fig. \ref{fig_6_c}
demonstrates that the strongest kick, $\eta =0.35{\pi /d}$, completely
destroys the soliton (recall the destruction never happens in the model with
the saturable nonlinearity). Therefore, there must exist critical values of
the kick separating the stable solitons, fuzzy beams, and complete
destruction. These borders are plotted in the top panel of Fig. \ref{fig_6_d}%
, in the range of powers $50<P<100$ (the average width of the static soliton
in the same region is plotted, versus $P$, in the bottom panel). Thus, the
large-amplitude solitons in the INPhC model with the cubic self-focusing are
not mobile at all. On the other hand, for dynamics of small-amplitude
solitons in the models with the saturable and cubic nonlinearities are
similar, due to the obvious expansion, $\left( 1+|u|^{2}\right) ^{-1}\approx
1-|u|^{2}$ (results for this case are not shown here in detail, as they are
less interesting).

\subsection{Collisions between moving quasi-compactons}

\begin{figure}[tbp]
\centering
\subfigure[]{ \label{fig_7_a}
\includegraphics[scale=0.3]{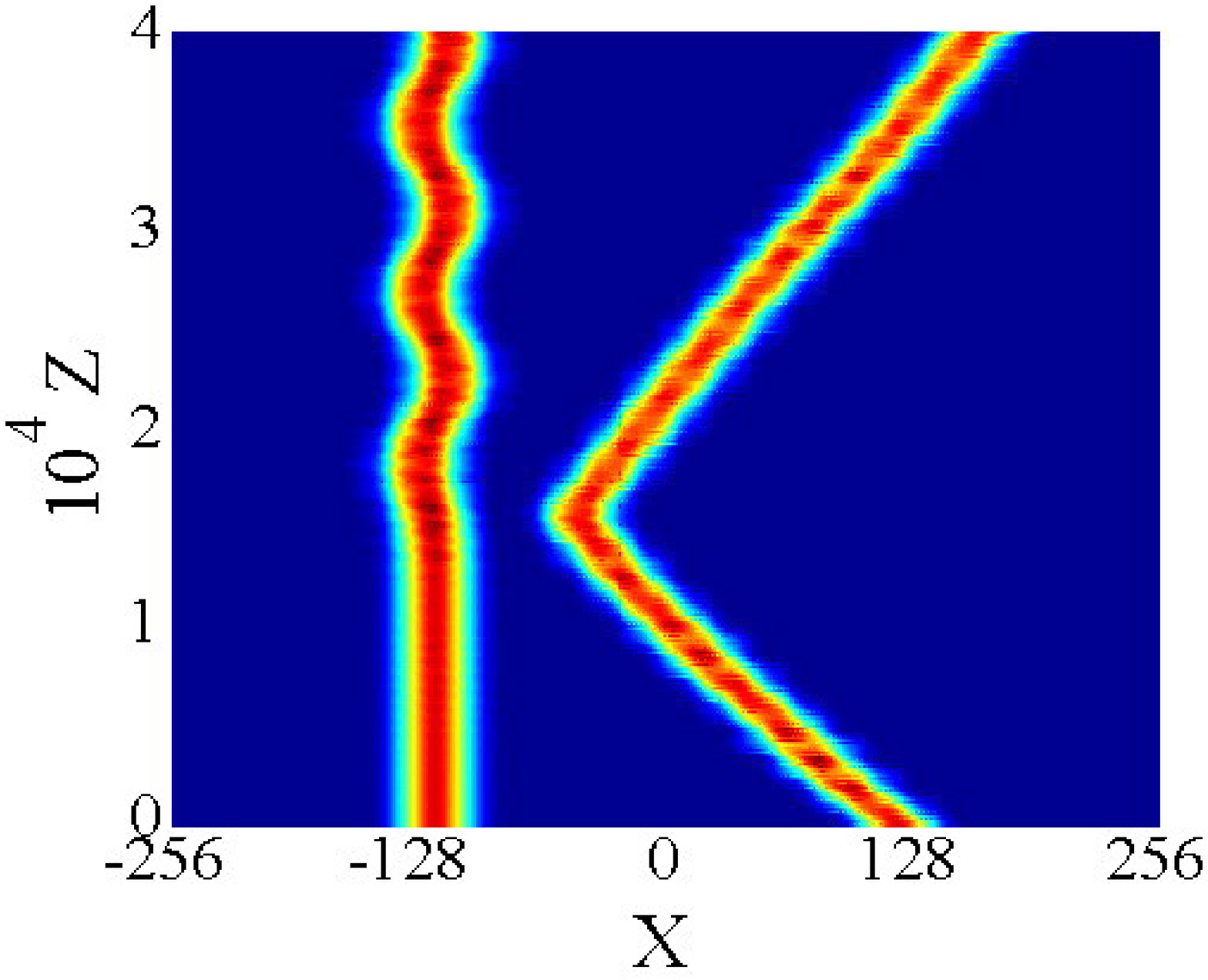}} \hspace{0.00in}%
\subfigure[]{ \label{fig_7_b}
\includegraphics[scale=0.3]{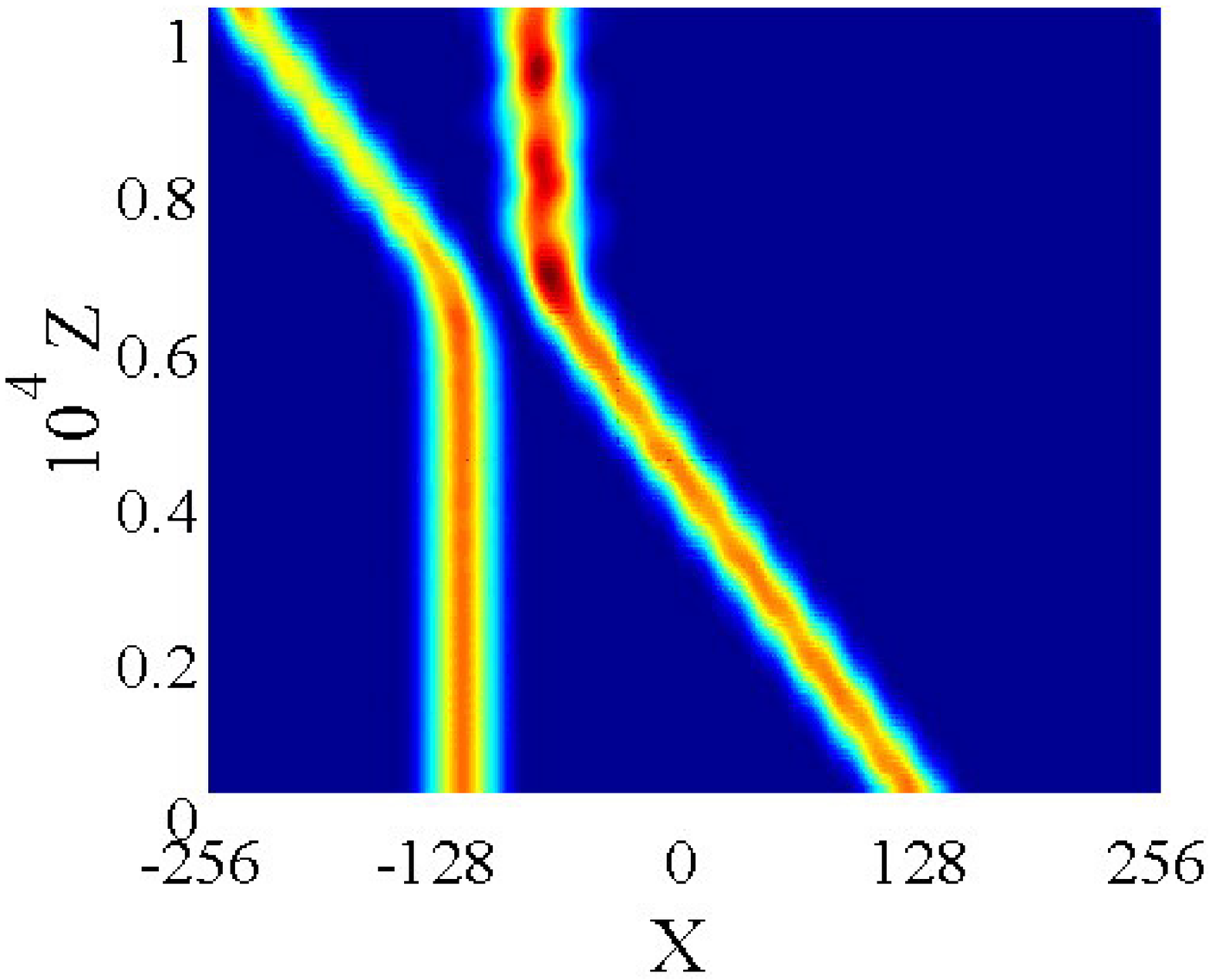}} \hspace{0.00in}%
\subfigure[]{ \label{fig_7_c}
\includegraphics[scale=0.3]{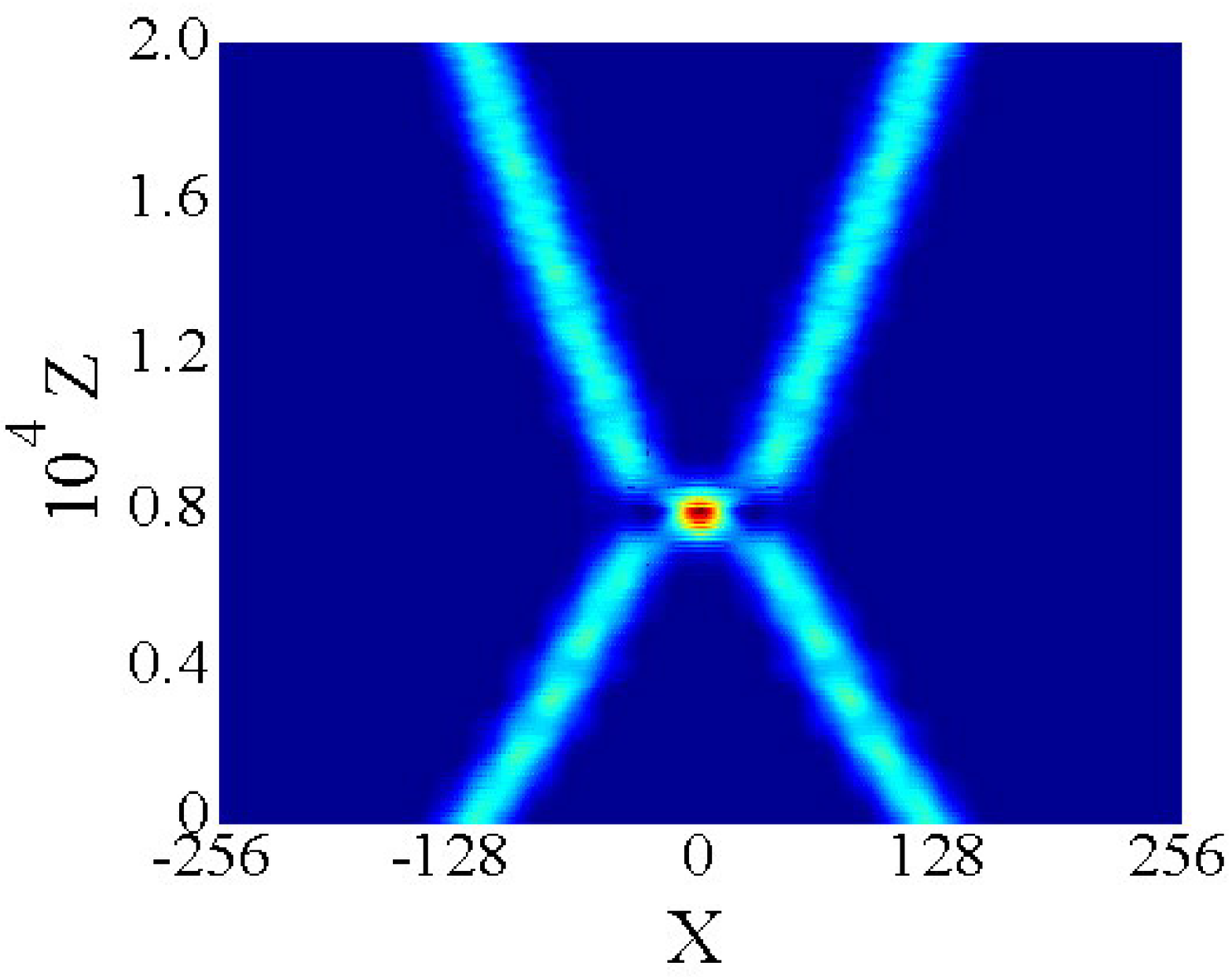}} \hspace{0.00in}%
\subfigure[]{ \label{fig_7_d}
\includegraphics[scale=0.3]{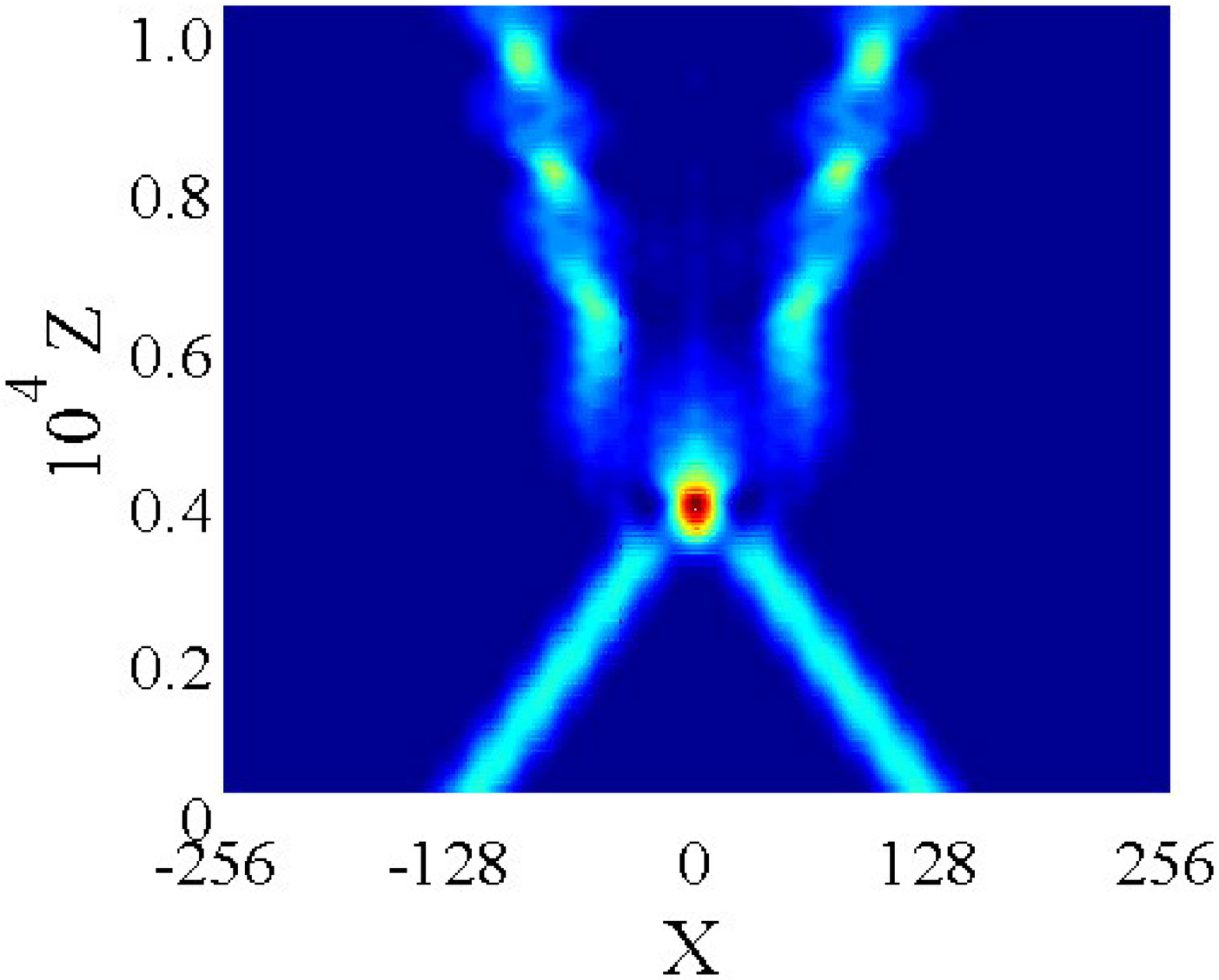}}
\caption{(Color online) Direct simulation of the collisions initiated by
kicks $\protect\eta _{1}=0$, $\protect\eta _{2}=0.1(\protect\pi /d)$ (a); $%
\protect\eta _{1}=0$, $\protect\eta _{2}=0.2(\protect\pi /d)$ (b); $\protect%
\eta _{1}=0.1(\protect\pi /d)$, $\protect\eta _{2}=0.1(\protect\pi /d)$ (c);
$\protect\eta _{1}=0.2(\protect\pi /d)$, $\protect\eta _{2}=0.2(\protect\pi %
/d)$ (d). Other parameters are $P=400$, $x_{0}=120$, and $V_{0}=0.02$. As in
the rest of the paper, the scaling is fixed by $d\equiv \allowbreak 20$.}
\label{fig_7}
\end{figure}

Collisions is a natural way to test interactions between solitons. Here we
present some numerical results for collisions between QCs in the INPhC with
the saturable nonlinearity. The initial state at $z=0$ is taken as a set of
two far separated kicked solitons:
\begin{equation}
u(x)=U(x+x_{0},P)e^{i\eta _{1}(x+x_{0})}+U(x-x_{0},P)e^{-i\eta
_{2}(x-x_{0})},  \label{init}
\end{equation}%
where $U(x\pm x_{0},P)$ are the QC pulses with power $P$, which are centered
at $x=\mp x_{0}$, and $\eta _{1,2}$ are the kicks applied to them.
Simulations of Eq. (\ref{NLSP}) with initial conditions (\ref{init}) are
displayed in Fig. \ref{fig_7}.

The simulations demonstrate the repulsive interaction between the colliding
QC. In Figs. \ref{fig_7_a} and \ref{fig_7_b}, they bounce back without
overlapping, because the collision velocity is small (nevertheless, one of
the solitons suffers a conspicuous perturbation in Fig. \ref{fig_7_a}). In
Figs. \ref{fig_7_c} and \ref{fig_7_d}, the larger velocities give rise to
the overlap between the colliding QCs, which results in the strong
inelasticity of the collision in the case displayed in Fig. \ref{fig_7_d}.

\section{Conclusions}

The objective of this work was to consider 1D solitons in the INPhC
(inverted nonlinear photonic crystal) with competing LL and NL
(linear and nonlinear lattices). Unlike recently studied models with
the cubic nonlinearity, we here consider the saturable
self-focusing. The 1D crystal of this type was recently fabricated,
using the SU-8 polymer material periodically doped with Rhodamine B,
which lends the medium the saturable nonlinearity. Combining
numerical methods and analytical approximations, we have
demonstrated that broad solitons are sharply localized in this
setting, thus taking the shape of the QCs (quasi-compactons). With
the increase of the total power, $P$, the QC remains a stable
object, which switches its position between the linear and nonlinear
layers, which form the INPhC. The width of the soliton increases
with $P$, which is a manifestation of the saturable nonlinearity.
The large width of the QC makes it a mobile object, unlike solitons
in the INPhC with the cubic nonlinearity. The threshold value of the
transverse kick (phase tilt), which is necessary to set the QC in
motion, was found as a function of $P$. The threshold value
gradually decays with the increase of $P$. Collisions between two
moving QCs were are also studied, by means of direct simulations.

The results reported in this work may be applied to the design of
all-optical data-processing schemes. In particular, the power-controlled
switch of the spatial soliton between adjacent layers, as well as the high
mobility of these solitons, may be quite relevant properties, in this
context.

This work may be extended in other directions. It particular, 2D
waveguide arrays, based on the RhB-SU-8 mixtures, can be fabricated
\cite{MNF,LJT}, which suggests to consider spatial solitons in
two-dimensional INPhC with the saturable nonlinearity.

Y.L. appreciates useful discussions with Dr. M. Feng.
This work was supported by the Chinese agencies NKBRSF (grant No.
G2010CB923204), NSFC(grant No. 11104083) and CNNSF(grant No. 10934011).%
%
%

%

\bibliographystyle{plain}
\bibliography{apssamp}

\end{document}